\documentclass[range]{ar2e}

\newcommand{\be}{\begin{equation}}
\newcommand{\ee}{\end{equation}}
\newcommand{\ba}{\begin{eqnarray}}
\newcommand{\ea}{\end{eqnarray}}
\newcommand{\ban}{\begin{eqnarray*}}
\newcommand{\ean}{\end{eqnarray*}}

\begin{document}

\input epsf.tex   

\input psfig.sty

\jname{Annu. Rev. Nucl. Particle Science}
\jyear{2007}
\jvol{57}

\title{What Do Electromagnetic Plasmas Tell Us about Quark-Gluon Plasma?}

\markboth{Mr\' owczy\' nski \& Thoma}{Real \& Quark-Gluon Plasmas}

\author{Stanis\l aw Mr\' owczy\' nski
\affiliation{Institute of Physics, \'Swi\c etokrzyska Academy, \\
ul. \'Swi\c etokrzyska 15, PL - 25-406 Kielce, Poland \\
and So\l tan Institute for Nuclear Studies, \\
ul. Ho\.za 69, PL - 00-681 Warsaw, Poland \\
e-mail: mrow@fuw.edu.pl}
Markus H. Thoma
\affiliation{Max-Planck-Institute for Extraterrestrial Physics, \\
P.O. Box 1312, D - 85741 Garching, Germany \\
e-mail: mthoma@mpe.mpg.de}}


\begin{abstract}

Since the quark-gluon plasma (QGP) reveals some obvious similarities to the 
well-known electromagnetic plasma (EMP), an accumulated knowledge on 
EMP can be used in the QGP studies. After discussing similarities and 
differences of the two systems, we present theoretical tools which are
used to describe the plasmas. The tools include: kinetic theory,
hydrodynamic approach and diagrammatic perturbative methods. We consider 
collective phenomena in the plasma with a particular emphasis on 
instabilities which crucially influence temporal evolution of the 
system. Finally, properties of strongly coupled plasma are discussed.

\end{abstract}

\maketitle

                                                                                
\section{INTRODUCTION}
                                                                                

The plasma - the ionized gas of electrons and ions - has been 
actively studied since its discovery in a discharge tube at the 
end of 19-th century. The term {\it plasma} was introduced 
by Irving Langmuir in 1929. Prospects to get a practically 
unlimited source of energy due to nuclear fusion reactions in a hot 
ionized gas of hydrogen isotopes, have stimulated a large scale 
program to study plasmas in terrestrial experiments for over half
a century. Plasmas are also actively studied by astrophysicists as it
appears the most common phase of matter. About 99\% of the entire 
visible Universe is in the plasma phase, not only stars are formed 
of ionized gas but the interstellar and intergalactic medium is also 
a plasma, although a very sparse one. Principles of the plasma 
physics can be found in e.g. the well-known textbooks 
\cite{Kra73,Ichimaru_1973}.

The quark-gluon plasma (QGP) is the system of quarks and gluons 
which are not confined in hadron's interiors but can freely move 
in a whole volume occupied by the system. A broad presentation
of the whole field of QGP physics is contained in three volumes of 
review articles \cite{QGP-Hwa1,QGP-Hwa2,QGP-Hwa3}; the lectures \cite{Mrowczynski:1998jr} can serve as an elementary introduction. 
Active studies of QGP started in the mid 1980s when relativistic 
heavy-ion collisions offered an opportunity to create a drop of QGP 
in a laboratory. The experimental programs at CERN and BNL provided 
an evidence of the QGP production at the early stage of nucleus-nucleus
collisions, when the system is extremely hot and dense, but properties 
of QGP remain enigmatic. So, one can ask: what do electromagnetic 
plasmas (EMP) tells us about QGP? 
  
The QGP reveals some obvious similarities to the well-known 
electromagnetic plasma, as Quantum Chromodynamics (QCD) describing 
the interactions of the quarks and gluons resembles Quantum 
Electrodynamics (QED) which governs interactions of charged
objects. Thus, some lessons from EMP should be useful in the 
exploration of QGP. The aim of this article  is to discuss what 
QGP physicists can actually learn from their EMP colleagues and how 
the huge accumulated knowledge on EMP can be used in the QGP studies. 
However, we have to be aware not only of similarities but of 
important differences between EMP and QGP. Some differences are 
of rather trivial origin but some are deeply rooted in dynamical 
foundations of the two systems.

Let us enumerate these `trivial' dissimilarities. QGP is usually 
relativistic or even ultrarelativistic while EMP is mostly 
nonrelativistic in laboratory experiments. The differences between 
the non-relativistic and relativistic plasmas go far beyond the 
kinematics of motion of plasma particles. For example, let us 
consider the plasma's composition. In the non-relativistic system, 
there are particles but no antiparticles and the particle's number 
is conserved. In the relativistic system we have both particles andantiparticles, as electrons and positrons in EMP, and the lepton
number - not the particle's number - is conserved. Particle number
density is not a right quantity to characterize the system. For 
this reason, the QGP physicists use the baryon and strangeness
densities.

Another `trivial' but very important distinctive feature of EMP 
is the huge mass difference between electrons and ions which is 
responsible for a specific dynamic role of heavy ions. The ions 
are usually treated as a passive background, which merely 
compensates the charge of electrons, but electro-ion collisions 
drive the system towards equilibrium and maintain the equilibrium. 
However, the energy transfer between electrons and ions is very 
inefficient and their mutual equilibration is very slow. Therefore,
we have electron and ion fluids of different temperatures for 
a relatively long time. There is nothing similar in QGP. There 
are heavy quarks - charm, bottom and top - which are, however, much 
less populated than the light quarks and gluons and their lifetime 
is short. Therefore, the heavy quarks hardly influence the QGP 
dynamics. 

The electromagnetic plasma, which is the closest analogue of QGP, 
is the relativistic system of electrons, positrons and photons. 
Such a plasma is actually studied in context of some astrophysical 
applications, e.g. supernovae explosions. Then, the differences between 
QGP and EMP are of dynamical origin - the first one is governed by 
Quantum Chromodynamics while the second one by Quantum Electrodynamics. 
The former theory is Abelian while the later one is nonAbelian with 
a prominent role of gluons which carry color charges and thus, not only 
mediate the interaction among colored quarks and antiquarks but interact 
among themselves. Gluons, in contrast to photons, also contribute to the 
density of color charges and to the color current. 

The most important common feature of EMP and QGP is the collective 
character of the dynamics. The range of electrostatic interaction 
is, in spite of the screening, usually much larger than the 
inter-particle spacing. There are many particles in the Debye 
sphere - the sphere of the radius equals the effective interaction 
range, and motion of these particles is highly correlated. There 
is a similar situation in the deconfined perturbative phase of QCD
\cite{Blaizot:2001nr}. The Debye mass is of order $gT$ where $g$ 
is the QCD constant and $T$ is the temperature. Since the particle 
density in QGP is of order $T^3$, the number of partons in the Debye 
sphere, which is roughly $1/g^3$, is large in the weakly coupled 
($1/g \gg 1$) QGP.

In various laboratory experiments, EMP is embedded in an external 
electromagnetic field. For example, the magnetic field is used
to trap the plasma, and there are numerous fascinating phenomena
occurring in such a situation. In the case of QGP produced in 
relativistic heavy-ion collisions, it is hard to imagine any 
external chromodynamic field applied to the plasma. Therefore, 
we will consider here only the systems where fields are 
self-consistently generated in the plasma.

Our article is organized as follows. Theoretical tools, which are 
used to describe the plasmas, are presented in Chapter~\ref{sec-tools}. 
The tools include: the kinetic theory, the hydrodynamic approach and 
diagrammatic methods of field theory. In Chapter~\ref{sec-collective} we 
discuss collective phenomena which are the most characteristic feature 
of plasmas. After explaining the phenomenon of screening, quasi-particle 
modes in the equilibrium and non-equilibrium plasma are presented. We pay 
much attention to instabilities which crucially influence plasma dynamics. 
The problem of particle's energy loss in a plasma is also discussed. 
Chapter~\ref{sec-strong} is devoted to the strongly coupled plasma which 
reveals particularly interesting properties.

Throughout the paper we use the natural units with $c=\hbar=k_B =1$
and the metric $(1,-1,-1,-1)$. However, there is a little complication 
here. Plasma physicists usually use the Gauss (CGS) units, where the 
fine structure constant equals $\alpha = e^2 \approx 1/137$, while 
the electromagnetic counterpart of the units usually applied in QCD 
is the so-called Heaviside-Lorentz system where the $4\pi$ factor 
does not show up in the Maxwell equations but $\alpha = e^2 /4 \pi$. 
We stick to the traditionally used units in the two fields of physics, 
and thus the factor of $4\pi$ has to be additionally taken into account 
to compare EMP to QGP formulas.


\section{THEORETICAL TOOLS}
\label{sec-tools}
                                                                                

\subsection{Transport theory}
\label{sec-kin}

Transport theory provides a natural framework to study equilibrium
and nonequilibrium plasmas. The central object of the theory is the 
distribution function which describes a time dependent distribution  
of particles in a phase-space spanned by the particle's momenta and 
positions. The distribution function of each plasma component evolves
due to the inter-particle collisions and the interaction with an
external and/or self-consistently generated mean-field. The two 
dynamical effects give rise to the collision and mean-field terms of 
a transport equation satisfied by the distribution function. 

\subsubsection{ELECTROMAGNETIC PLASMA}

A formulation of the kinetic theory of relativistic plasma can
be found in \cite{Gro80}. The distribution function is denoted 
as $f_n({\bf p},x)$ with the index $n$ labeling plasma components: 
electrons, posi\-trons, ions. Spin is usually treated as an internal 
degree of freedom. The function depends on the four-position 
$x\equiv(t,{\bf x})$ and the three-momentum ${\bf p}$. The four-momentum 
$p$ obeys the mass-shell constraint $p^2 = m^2$, where $m$ is the 
particle mass, and then $p\equiv(E_p,{\bf p})$ with 
$E_p \equiv \sqrt{m^2 + {\bf p}^2}$.

The distribution function satisfies the transport equation
\be
\label{transport}
\Big( p_\mu \partial^\mu + q_n p^\mu F_{\mu \nu} \partial^\nu_p \Big)
f_n({\bf p},x) = C[f_n] \;,
\ee
where $C[f_n]$ denotes the collision term, $q_n$ is the charge of the 
plasma species $n$ and $F^{\mu \nu}$ is the electromagnetic strength 
tensor which either represents an external field applied to the system 
or/and is generated self-consistently by the four-currents present in 
the plasma
$$
\partial^\mu F_{\mu \nu} = 4\pi j_\nu \;,
$$ 
where
\be
\label{current}
j^\mu(x) = \sum_n q_n \int \frac{d^3p}{(2\pi)^3} 
\frac{p^\mu}{E_p} \, f_n({\bf p},x) \;.
\ee

The transport equation can be solved in the linear response 
approximation. The equation is linearized around the stationary 
and homogeneous state described by the distribution 
$\bar f_n({\bf p})$. The state is also assumed to be neutral and 
there are no currents. The distribution function is then decomposed 
as
$$
f_n ({\bf p},x) = \bar f_n ({\bf p})  + \delta f_n({\bf p},x) \;,
$$
where $\bar f_n ({\bf p}) \gg \delta f_n ({\bf p},x)$.

The transport equation linearized in $\delta f_n$ and $F^{\mu \nu}$ 
can be exactly solved after the Fourier transformation, which is
defined as
\be
\label{Fourier}
f(k) = \int d^4x \: e^{ikx} f(x) \;,\;\;\;\;\;\;
f(x) = \int \frac{d^4k}{(2\pi)^4} \: e^{-ikx} f(k) \;.
\ee
Then, one finds $\delta f_n({\bf p},k)$, which is the Fourier 
transform of $\delta f_n({\bf p},x)$, and the induced current, 
which can be written as 
\be
\label{Pi-def}
\delta j^\mu(k) = -\Pi^{\mu \nu}(k) A_\nu(k) \;,
\ee
with the polarization tensor equal to
\be
\label{Pi-kinetic}
\Pi^{\mu \nu}(k) = 4\pi \sum_n q_n^2 
\int \frac{d^3p}{(2\pi)^3} \bar f_n ({\bf p}) \;
{(p\cdot k)^2 g^{\mu\nu} + k^2 p^{\mu} p^{\nu}
-(p\cdot k)(k^\mu p^\nu + k^\nu p^\mu) 
\over(p\cdot k)^2} \;.
\ee
The tensor is symmetric ($\Pi^{\mu \nu}(k)=\Pi^{\nu \mu}(k)$)
and transverse ($k_\mu \Pi^{\mu \nu}(k) = 0$) which guarantees 
that the current given by Equation~\ref{Pi-def} is gauge 
independent. 

For isotropic plasmas, the polarization tensor has only 
two independent components which are usually chosen as 
\ba
\Pi _L(k) & = & \Pi_{00}(k)\; ,\nonumber \\[1mm]
\Pi _T(k) & = & \frac {1}{2}\> \left (\delta _{ij} 
- \frac {k_ik_j}{{\bf k}^2}
\right )\> \Pi _{ij}(k)\; ,
\label{pol-L-T}
\ea
where the indices $i,j = 1,2,3$ label three-vector and tensor 
components. In the case of an ultrarelativistic ($T\gg m$) 
electron-positron equilibrium plasma, the momentum integral in 
Equation~\ref{Pi-kinetic} can be performed analytically 
in the high-temperature limit ($T\gg \omega,|{\bf k}|$), and the 
result derived already in 1960 by Silin \cite{Silin_1960} reads
\ba
\Pi _L(k) & = & -3\> m_\gamma ^2\> 
\left[ 1-\frac{\omega}{2|{\bf k}|} \> 
\ln \frac{\omega +|{\bf k}|}{\omega -|{\bf k}|}\right] \; ,
\nonumber \\[1mm]
\Pi _T(k) & = & \frac {3}{2}\> m_\gamma ^2\> \frac {\omega^2}{{\bf k}^2}\>
\left [1-\left (1-\frac{{\bf k}^2}{\omega^2}\right )\> 
\frac{\omega}{2|{\bf k}|}\> 
\ln \frac{\omega +|{\bf k}|}{\omega -|{\bf k}|} \right ]\;,
\label{polar-equi}
\ea
where $k\equiv (\omega,{\bf k})$ and $m_\gamma \equiv eT/3$ denotes 
the {\it thermal photon mass} generated by the interaction of the 
photons with the electrons and positrons.

The above polarization tensor was found in the collisionless
limit of the transport equation. The effect of collisions can be easily
taken into account if the so-called BGK collision term is used in the 
transport equation \cite{Alexandrov_1984}. The result for an
ultrarelativistic equilibrium plasma is given in \cite{Carrington_2004}. 

\subsubsection{QUARK-GLUON PLASMA}

The transport theory of QGP \cite{Elze:1989un,Mrowczynski:np}
appears to be much more complicated than its electromagnetic 
counterpart. The distribution function of quarks $Q({\bf p},x)$ is 
a hermitian $N_c\times N_c$ matrix in  color space (for a SU($N_c$) 
color gauge group). The distribution function is gauge dependent and 
it transforms under a local gauge transformation $U(x)$ as
\be
\label{Q-transform}
Q({\bf p},x) \rightarrow U(x) \, Q({\bf p},x) \, U^{\dag }(x) \;.
\ee
Here and in the most cases below, the color indices are suppressed.
The distribution function of antiquarks, which we denote by 
$\tilde Q({\bf p},x)$, is also a hermitian $N_c\times N_c$ matrix 
and it transforms according to Equation~\ref{Q-transform}. The 
distribution function of gluons is a hermitian 
$(N_c^2-1)\times (N_c^2-1)$ matrix which transforms as
\be
\label{G-transform}
G({\bf p},x) \rightarrow {\cal U}(x) \: G({\bf p},x) 
\:{\cal U}^{\dag }(x) \;,
\ee
where
$$
{\cal U}_{ab}(x) = 2{\rm Tr}\bigr[\tau^a U(x) 
\tau^b U^{\dag }(x)] \ ,
$$
with $\tau^a ,\; a = 1,...,N_c^2-1$ being the SU($N_c$) group
generators in the fundamental representation with 
${\rm Tr} (\tau^a \tau^b) = \frac12 \delta^{ab}$. 

The color current is expressed in the fundamental representation as
\ba
\label{col-current}
j^{\mu }(x) = &-&\frac{g}{2} \int \frac{d^3p}{(2\pi)^3} \; p^\mu \;
\Big[ Q({\bf p},x) - \tilde Q ({\bf p},x)
\\ \nonumber
&-& {1 \over N_c}{\rm Tr}\big[Q({\bf p},x) - \tilde Q ({\bf p},x)\big] 
+ 2 \tau^a {\rm Tr}\big[T^a G({\bf p},x) \big]\Big] \;,
\ea
where $g$ is the QCD coupling constant. A sum over helicities, 
two per particle, and over quark flavors $N_f$ is understood in 
Equation~\ref{col-current}, even though it is not explicitly written
down. The SU($N_c$) generators in the adjoint representation are
expressed through the structure constants  $T^a_{bc} = -i f_{abc}$,
and are normalized as ${\rm Tr}[T^aT^b]= N_c \delta^{ab}$. The
current can be decomposed as $j^\mu (x) = j^\mu_a (x) \tau^a$ 
with $j^\mu_a (x) = 2 {\rm Tr} (\tau_a j^\mu (x))$. The 
distribution functions, which are proportional to the unit matrix 
in color space, are gauge independent and they provide the color 
current (Equation~\ref{col-current}) which identically vanishes.

Gauge invariant quantities are given by the traces of the
distribution functions. Thus, the baryon current and the 
energy-momentum tensor read
\ban
b^{\mu }(x) &=& {1 \over 3} \int \frac{d^3p}{(2\pi)^3}\; p^{\mu} \;
{\rm Tr}\Big[ Q({\bf p},x) - \tilde Q ({\bf p},x) \Big] \; , 
\\[2mm]
T^{\mu \nu}(x) &=& 
\int \frac{d^3p}{(2\pi)^3}\; p^{\mu} p^{\nu} \;
{\rm Tr}\Big[ Q({\bf p},x) 
+ \tilde Q({\bf p},x) + G({\bf p},x) \Big] \; ,
\ean
where we use the same symbol ${\rm Tr}[\cdots]$ for the trace 
both in the fundamental and adjoint representations.

The distribution functions of quarks, antiquarks and gluons 
satisfy the transport equations:
\ba
p^{\mu} D_{\mu}Q({\bf p},x) + {g \over 2}\: p^{\mu}
\left\{ F_{\mu \nu}(x), \partial^\nu_p Q({\bf p},x) \right\}
&=& C[Q,\tilde Q,G] \;,
\label{transport-q}  \\ [2mm]
p^{\mu} D_{\mu}\tilde Q({\bf p},x) - {g \over 2} \: p^{\mu}
\left\{ F_{\mu \nu}(x), \partial^\nu_p \tilde Q({\bf p},x)\right\}
&=& \tilde C[Q,\tilde Q,G]\;,
\label{transport-barq} \\ [2mm]
p^{\mu} {\cal D}_{\mu}G({\bf p},x) + {g \over 2} \: p^{\mu}
\left\{ {\cal F}_{\mu \nu}(x), \partial^\nu_p G({\bf p},x) \right\}
&=& C_g[Q,\tilde Q,G]\;,
\label{transport-gluon}
\ea
where $\{...,...\}$ denotes the anticommutator and $\partial^\nu_p$ 
the four-momentum derivative\footnote{As the distribution functions
do not depend on $p_0$, the derivative over $p_0$ is identically zero.}; 
the covariant derivatives $D_{\mu}$ and ${\cal D}_{\mu}$ act as
$$
D_{\mu} = \partial_{\mu} - ig[A_{\mu}(x),...\; ]\;,\;\;\;\;\;\;\;
{\cal D}_{\mu} = \partial_{\mu} - ig[{\cal A}_{\mu}(x),...\;]\;,
$$
with $A_{\mu }$ and ${\cal A}_{\mu }$ being four-potentials
in the fundamental and adjoint representations, respectively:
$$
A^{\mu }(x) = A^{\mu }_a (x) \tau^a \;,\;\;\;\;\;
{\cal A}^{\mu }(x) = T^a A^{\mu }_a (x) \; .
$$
The strength tensor in the fundamental representation is
$F_{\mu\nu}=\partial_{\mu}A_{\nu} - \partial_{\nu}A_{\mu}
-ig [A_{\mu},A_{\nu}]$, while  ${\cal F}_{\mu \nu}$ denotes the 
field strength tensor in the adjoint representation. $C, \tilde C$
and $C_g$ represent the collision terms.

The transport equations are supplemented by the Yang-Mills equation 
describing generation of the gauge field
\be
\label{yang-mills}
D_{\mu} F^{\mu \nu}(x) = j^{\nu}(x)\; ,
\ee
where the color current is given by Equation~\ref{col-current}.

As in the case of the EM plasma, the transport equations, which are 
linearized around a stationary, homogeneous and colorless state can 
be solved. Because of the color neutrality assumption, the analysis 
is rather similar to that of the Abelian plasma, and it ends up with 
the polarization tensor which is proportional to the unit matrix in 
the color space and has the form of Equation~\ref{Pi-kinetic}. 

As in the case of EMP, the collisions can be easily taken into 
account using the approximate BGK collision terms \cite{Manuel:2004gk,Schenke:2006xu}. Within a more realistic 
approach color charges are treated in a similar way as spin degrees 
of freedom, and ones uses the so-called Waldmann-Snider collision 
terms \cite{Arnold:1998cy,Manuel:2003zr} which are usually applied 
to study spin transport.

\subsection{Hydrodynamic approach}

Within the hydrodynamic approach, the plasma is treated as a liquid
and it is described in terms of macroscopic variables which obey
the equations of motion resulting from the conservation laws. 
The fluid equations are applied to a large variety of plasma
phenomena but, depending of the time scale of interest, the actual
physical content of the equations is rather different.

Real hydrodynamics deals with systems which are in local equilibrium, 
and thus it is only applicable at sufficiently long time scales. 
The continuity and the Euler or Navier-Stokes equations are 
supplemented by the equation of state to form a complete set 
of equations. The equations can be derived from  kinetic theory, 
using the distribution function of local equilibrium which by 
definition maximizes the entropy density, and thus, it cancels the 
collision terms of the transport equations. 

In the electron-ion plasma there are several time scales of
equilibration. The electron component of the plasma reaches
the equilibrium in the shortest time, then ions are equilibrated 
but for a relatively long time the electron and ion temperatures 
remain different from each other, as the energy transfer between 
electrons to ions is rather inefficient. This happens due to 
the huge mass difference between electrons and ions. 

When the electrons have reached local equilibrium with their own 
temperature and hydrodynamic velocity, the collision terms of the kinetic 
equations representing electron-electron collisions vanish while the 
collision terms due to electron-ion collisions can be neglected as 
they influence the electron distribution function only at a sufficiently 
long time scale. Then, one obtains hydrodynamic equations of an 
electron fluid. When the ion component is also equilibrated we have 
two fluids with different temperatures and hydrodynamic velocities. 
At the times scales when the fluid equations are applicable, the plasma 
can be treated as locally neutral. Charge fluctuations are obviously 
possible but they disappear fast as the electric field generated by 
the local charges induces the currents which in turn neutralize the 
charges. Since the plasma is nearly an ideal conductor, the process 
of plasma neutralization is very fast. Due to the charge neutrality 
of the plasma, the electric field is not present in the fluid equations 
and we end up with the {\it magnetohydrodynamics} where the pressure
gradients and magnetic field drive the plasma dynamics.

As explained above, the regime of magnetohydrodynamics appears because
there is a heavy positive component of the plasma (ions) and a light 
negative component (electrons). There is no QCD analogue of 
magnetohydrodynamics as every quark or gluon can carry opposite color 
charges. Therefore, when local equilibrium is reached various color 
components of the plasma have the same temperatures and hydrodynamic 
velocities \cite{Manuel:2003zr}. Since the quark-gluon system 
becomes color neutral even before the local equilibration is reached 
\cite{Arnold:1998cy,Manuel:2004gk}, we deal with hydrodynamics 
of neutral fluid where the chromodynamic fields are absent. Such
a relativistic hydrodynamics of colorless QGP has been actively 
studied over the last two decades \cite{Kolb:2003dz,Huovinen:2006jp}. 

The hydrodynamic equations, which actually express macroscopic 
conservation laws, hold not only for systems in local equilibrium 
but for systems out of equilibrium as well. The equations can be 
then applied at time scales significantly shorter than that of local 
equilibration. At such a short time scale, the collision terms of 
the transport equations can be neglected entirely. However, extra 
assumptions are then needed to close the set of equations, as 
the (equilibrium) equation of state cannot be used. Plasma 
physicists developed several methods to close the set of equations,
and thus fluid equations are used to study bulk features of short 
time scale phenomena in the plasmas. To get more detailed information, 
the kinetic theory is needed. Since the fluid equations are noticeably 
simpler than the kinetic ones, the hydrodynamic approach is frequently 
used in numerical simulations of plasma evolution, studies of nonlinear 
dynamics, etc. 

Below we derive the fluid equations for EMP and QGP from the 
respective kinetic theory. Since the fluid approach under consideration
is supposed to hold at sufficiently short time scales, we use the 
collisionless transport equations.

\subsubsection{ELECTROMAGNETIC PLASMA}

We assume here that there are several streams in the relativistic 
plasma system and that the distribution functions of each plasma 
component (electrons, positrons, ions) belonging of each stream 
satisfy the collisionless transport equation. The equations are 
coupled only through the electromagnetic mean field which is 
generated by the current coming from all streams. The field in 
turn interacts with every stream. 

Integrating the collisionless transport Equation~\ref{transport}
over momentum, one finds the continuity equation
\be
\label{cont-eq}
\partial_\mu n^\mu_\alpha = 0 \;,
\ee
where the four-flow is
\be
\label{flow}
n^\mu_\alpha (x) \equiv \int \frac{d^3p}{(2\pi)^3} \;
p^\mu f_\alpha({\bf p},x) \;.
\ee
The index $\alpha$ labels simultaneously the streams and plasma 
components.

Multiplying the transport Equation~\ref{transport} by the 
four-momentum $p$ and integrating it over momentum, we get
\be
\label{en-mom-eq}
\partial_\mu T^{\mu \nu}_\alpha+ q_\alpha n^\mu_\alpha  
F_{\mu}^{\;\; \nu}= 0 \;,
\ee
where the energy-momentum tensor is
\be
\label{en-mom}
T^{\mu \nu}_\alpha (x) \equiv \int \frac{d^3p}{(2\pi)^3}\;
p^\mu p^\nu  f_\alpha({\bf p},x) \;.
\ee

The structure of $n^\mu_\alpha$ and  $T^{\mu \nu}_\alpha$ is 
assumed to be that of the ideal fluid in local thermodynamic 
equilibrium. Thus, one has
\ba
\label{flow-id}
n^\mu_\alpha(x)  &=& n_\alpha (x) \, u_\alpha^\mu(x) \;, \\[2mm]
\label{en-mom-id}
T^{\mu \nu}_\alpha(x)  &=& \big[\epsilon_\alpha (x) + p_\alpha (x)\big]
u^\mu(x) u^\nu(x)
- p_\alpha (x) \, g^{\mu \nu}\;.
\ea

To obtain the relativistic version of the Euler equation,
Equation~\ref{en-mom-eq} needs to be manipulated following 
\cite{Lan63}. Substituting the energy-momentum tensor of the 
form of Equation~\ref{en-mom-id} into Equation~\ref{en-mom-eq} 
and projecting the result on the direction of $u^\mu_\alpha$, 
one finds
\be
\label{ener-den-eq}
u_{\alpha \nu} \partial_\mu T^{\mu \nu}_\alpha =
u^\mu_\alpha \partial_\mu \epsilon_\alpha +
(\epsilon_\alpha + p_\alpha ) \partial_\mu  u^\mu_\alpha = 0\;.
\ee
Computing $\partial_\mu T^{\mu \nu}_\alpha
- u_\alpha^\nu u_{\alpha \rho} \partial_\mu T^{\mu \rho}_\alpha$,
one gets the Lorentz covariant form of the Euler equation
\be
\label{cov-Euler}
M^\nu_\alpha \equiv(\epsilon_\alpha + p_\alpha )
u_{\alpha \mu} \partial^\mu  u^\nu_\alpha +
(u^\mu_\alpha u^\nu_\alpha \partial_\mu - \partial^\nu )p
- q_\alpha n_\alpha   u_{\alpha \mu} F^{\mu \nu}= 0\;.
\ee
The equation in a more familiar form is given by
${\bf M}_\alpha - {\bf v}_\alpha M^0_\alpha = 0$. Namely, 
\be
\label{Euler}
(\epsilon_\alpha + p_\alpha ) \gamma_\alpha^2
\Big( {\partial \over \partial t}
+ {\bf v}_\alpha \cdot \nabla \Big) {\bf v}_\alpha
+ \Big(\nabla + {\bf v}_\alpha {\partial \over \partial t} \Big)p_\alpha- q_\alpha n_\alpha \gamma_\alpha
\big[{\bf E} - {\bf v}_\alpha ({\bf v}_\alpha \cdot {\bf E} )
+ {\bf v}_\alpha \times {\bf B}\big] = 0 \;,
\ee
where the four-velocity $u_\alpha^\mu$ was expressed as 
$u_\alpha^\mu = (\gamma_\alpha, \gamma_\alpha {\bf v}_\alpha)$
with $\gamma_\alpha \equiv (1 - {\bf v}_\alpha^2)^{-1/2}$.

In the nonrelativistic limit (which is easily obtained when
the velocity of light $c$ is restored in the equation),
Equation~\ref{Euler} gets the well-known form
\be
\label{NR-Euler}
\Big( {\partial \over \partial t}+ {\bf v}_\alpha \cdot \nabla \Big) 
{\bf v}_\alpha
+ \frac{1}{m_\alpha n_\alpha }\nabla p_\alpha
- \frac{q_\alpha}{m_\alpha}
\big({\bf E} + {\bf v}_\alpha \times {\bf B}\big) = 0 \;.
\ee

The fluid Equations~\ref{cont-eq}, \ref{en-mom-eq} with 
$n^\mu_\alpha$ and $T^{\mu \nu}_\alpha$ given by 
Equations~\ref{flow-id}, \ref{en-mom-id} do not constitute a
closed set of equations - there are 5 equations and 6 
unknown functions: $n_\alpha$, $p_\alpha$, $\epsilon_\alpha$ 
and 3 components of $u^\mu_\alpha$ (because of the constraint
$u^\mu_\alpha u_{\mu \,\alpha}=1$, one component of 
$u^\mu_\alpha$ can be eliminated). There are several
methods to close the set of equations. In particular, assuming
that the system's dynamics is dominated by the mean-field
interaction, one can neglect the pressure gradients. One 
can also add an equation which relates $p_\alpha$ to 
$\epsilon_\alpha$. The relation is usually called the equation
of state, but one should be aware that the plasma system is 
not in equilibrium, and the thermodynamic relations in general 
do not hold.  

In the ultrarelativistic limit when the characteristic particle's
energy (the temperature of the equilibrium system) is much larger
than the particle's mass, and thus $p^2 = 0$, the energy-momentum 
tensor is traceless ($T^\mu_{\mu \, \alpha} = 0$), as follows from 
Equation~\ref{en-mom} for $p^2 = 0$. Then, Equation~\ref{en-mom-id} 
combined with the constraint $u_\alpha^\mu(x)u_{\alpha \,\mu}(x) = 1$ 
provides the desired relation
\be
\label{EoS}
\epsilon_\alpha (x) = 3 p_\alpha (x) \;,
\ee 
which coincides with the equation of state of an ideal gas of 
massless particles.

Since the distribution functions of every plasma component belonging
to every stream are assumed to obey the collisionless transport
equation, we have a separated set of fluid equations for every 
plasma component of every stream. The equations are coupled only
through the electromagnetic mean field. More precisely, the  
electrons, positrons and ions of every stream contribute to the
current generating the field which in turn interacts with the
streams. 

The fluid equations can be solved in the linear response
approximation. The equations are linearized around the stationary 
and homogeneous state described by $\bar n_\alpha$ and 
$\bar u^\mu_\alpha$. The state is neutral and there are no 
currents {\it i.e.}
\be
\label{neutral-cov}
\sum_\alpha  \bar n_\alpha \bar u^\mu_\alpha = 0 \;.
\ee
The charge density is decomposed as
\be
n_\alpha (x) = \bar n_\alpha  + \delta n_\alpha(x) \;,
\ee
where $\bar n_\alpha \gg \delta n_\alpha$. The fully analogous 
decomposition of the hydrodynamic velocity $u^\mu_\alpha$, pressure 
$p_\alpha$ and energy density $\epsilon_\alpha$ is also adopted.

The set of the continuity and Euler equations linearized
in $\delta n_\alpha$, $\delta u^\mu_\alpha$, $\delta p_\alpha$, 
$\delta \epsilon_\alpha$, and $F^{\mu \nu}$ can be exactly solved
after they are Fourier transformed. Thus, one finds 
$\delta n_\alpha(k)$, $\delta u^\mu_\alpha(k)$ when the set of 
fluid equations is closed by neglecting the pressure gradients. 
If the equation of state is used, one also finds 
$\delta \epsilon_\alpha(k)$.

Keeping in mind that the induced current equals
$$
\delta j^\mu = \sum_\alpha
(q_\alpha \bar n_\alpha \delta u^\mu_\alpha
+ q_\alpha \delta n_\alpha \, \bar u^\mu_\alpha )\;,
$$
one finds from Equation~\ref{Pi-def}
\ba
\nonumber 
\Pi^{\mu \nu}(k) &=& \sum_\alpha
\frac{4\pi q_\alpha^2 \bar n_\alpha^2}
{\bar \epsilon_\alpha + \bar p_\alpha} \:
\frac{1}{(\bar u_\alpha \cdot k)^2}
\Big[ k^2 \bar u^\mu_\alpha \bar u^\nu_\alpha 
+ (\bar u_\alpha \cdot k)^2 g^{\mu \nu}
- (\bar u_\alpha \cdot k)
(k^\mu \bar u^\nu_\alpha + k^\nu \bar u^\mu_\alpha )
\\[2mm] \label{Pi2}
&+& \frac{(\bar u_\alpha \cdot k)k^2
(k^\mu \bar u^\nu_\alpha + k^\nu \bar u^\mu_\alpha )
- (\bar u_\alpha \cdot k)^2 k^\mu k^\nu
- k^4  \bar u^\mu_\alpha \bar u^\nu_\alpha}
{k^2 + 2(\bar u_\alpha \cdot k)^2} \: \Big]\;.
\ea
The first term gives the polarization tensor when the pressure 
gradients are neglected while the second term gives the effect 
of the pressure gradients due to the equation of state given by Equation~\ref{EoS}. The first term is symmetric 
($\Pi^{\mu \nu}(k)=\Pi^{\nu \mu}(k)$) and transverse 
($k_\mu \Pi^{\mu \nu}(k) = 0$). The second term 
is symmetric and transverse as well. Thus, the whole polarization
tensor (Equation~\ref{Pi2}) is symmetric and transverse.
The first term of Equation~\ref{Pi2} can be obtained from
the kinetic theory result (Equation~\ref{Pi-kinetic}) with
the distribution function $\bar f_n({\bf p})$ proportional to 
$\delta^{(3)}({\bf p} - (\bar \epsilon_\alpha + \bar p_\alpha) 
{\bf u}_\alpha /\bar n_\alpha )$. Thus, the first term
neglects the thermal motion of plasma particles while the second
term takes the effect into account.

\subsubsection{QUARK-GLUON PLASMA}

The fluid approach presented here follows the formulation 
given in \cite{Manuel:2006hg}. As in the EMP case, we assume that 
there are several streams in the plasma system and that the 
distribution functions of quarks, antiquarks and gluons of each 
stream satisfy the collisionless transport equation. The streams 
are labeled with the index $\alpha$. 

Further analysis is limited to quarks but inclusion of anti-quarks 
and gluons is straightforward. The distribution function of quarks 
belonging to the stream $\alpha$ is denoted as $Q_\alpha({\bf p},x)$. 
Integrating over momentum the collisionless transport 
Equation~\ref{transport-q} satisfied by $Q_\alpha$, one finds the 
covariant continuity equation
\be
\label{cont-eq-qgp}
D_\mu n^\mu_\alpha = 0 \;,
\ee
where $n^\mu_\alpha$ is $N_c\times N_c$ matrix defined as
\be
\label{flow-qgp}
n^\mu_\alpha (x) \equiv \int \frac{d^3p}{(2\pi)^3}\;
p^\mu Q_\alpha({\bf p},x) \;.
\ee
The four-flow $n^\mu_\alpha$ transforms under gauge transformations
as the quark distribution function, {\it i.e.} according to
Equation~\ref{Q-transform}.

Multiplying the transport Equation~\ref{transport-q} by the 
four-momentum and integrating the product over momentum, we get
\be
\label{en-mom-eq-qgp}
D_\mu T^{\mu \nu}_\alpha
- {g \over 2}\{F_{\mu}^{\;\; \nu}, n^\mu_\alpha \}= 0 \;,
\ee
where the energy-momentum tensor is
\be
\label{en-mom-q-qgp}
T^{\mu \nu}_\alpha (x) \equiv \int \frac{d^3p}{(2\pi)^3}\;
p^\mu p^\nu  Q_\alpha({\bf p},x) \;.
\ee

We further assume that the structure of $n^\mu_\alpha$ and
$T^{\mu \nu}_\alpha$ is
\ba
\label{flow-id-qgp}
n^\mu_\alpha(x)  &=& n_\alpha (x) \, u_\alpha^\mu(x) \;,
\\[2mm]
\label{en-mom-id-qgp}
T^{\mu \nu}_\alpha(x)  &=& {1 \over 2}
\big(\epsilon_\alpha (x) + p_\alpha (x)\big)
\big\{u^\mu_\alpha (x), u^\nu_\alpha (x) \big\}
- p_\alpha (x) \, g^{\mu \nu}\;,
\ea
where the hydrodynamic velocity $u^\mu_\alpha$ is, as
$n_\alpha$, $\epsilon_\alpha$ and $p_\alpha$, a $N_c\times N_c$
matrix. The anticommutator of $u^\mu_\alpha$ and $u^\nu_\alpha$ is
present in Equation~\ref{en-mom-id-qgp} to guarantee the symmetry
of $T^{\mu \nu}_\alpha$ with respect to $\mu \leftrightarrow \nu$
which is evident in Equation~\ref{en-mom-q-qgp}.

In the case of an Abelian plasma, the relativistic version of the
Euler equation is obtained from Equation~\ref{en-mom-eq-qgp} by 
removing from it the part which is parallel to $u^\mu_\alpha$. An 
analogous procedure is not possible for the non-Abelian plasma 
because the matrices $n_\alpha$, $u^\mu_\alpha$, and $u^\nu_\alpha$, 
in general, do not commute with each other. Thus, one has to work directly with Equations~\ref{cont-eq-qgp}, \ref{en-mom-eq-qgp} 
with $n^\mu_\alpha$ and $T^{\mu \nu}_\alpha$ defined by 
Equations~\ref{flow-id-qgp}, \ref{en-mom-id-qgp}. The equations 
have to be supplemented by the Yang-Mills Equation~\ref{yang-mills} 
with the color current of the form 
\be
\label{hydro-current-qgp} 
j^\mu(x) = -\frac{g}{2} \sum_\alpha \Big(n_\alpha u^\mu_\alpha 
- {1 \over N_c}{\rm Tr}\big[n_\alpha u^\mu_\alpha  \big]\Big) \;, 
\ee 
where only the quark contribution is taken into account.

The fluid Equations~\ref{cont-eq-qgp}, \ref{en-mom-eq-qgp}, as 
their EM counter part, do not form a closed set of equations but 
can be closed analogously. The only difference is that the
equation of state (Equation~\ref{EoS}) relates to each other the 
matrix value functions $\epsilon_\alpha$ and $p_\alpha$.  

As in the case of the EM plasma, the fluid Equations~\ref{cont-eq}, 
\ref{en-mom-eq}, which are linearized around a stationary, homogeneous 
and colorless state described by $\bar n_\alpha$, $\bar \epsilon_\alpha$, 
$\bar p_\alpha$ and $\bar u^\mu_\alpha$, can be solved 
\cite{Manuel:2006hg}.  Because of the color neutrality assumption, 
$\bar n_\alpha$, $\bar \epsilon_\alpha$, $\bar p_\alpha$ and 
$\bar u^\mu_\alpha$ are all proportional to the unit matrix in the 
color space, the analysis is rather similar to that of the Abelian plasma, 
and one ends up with the polarization tensor from Equation~\ref{Pi2} which 
is proportional to the unit matrix in the color space.

\subsection{Diagrammatic methods}

Various characteristics of the weakly coupled plasma can be calculated 
using the perturbative expansion, that is diagrammatic methods of field 
theory. It requires a generalization of the Feynman rules applicable 
to processes, which occur in vacuum, to processes in many body plasma 
systems. When the plasma is in thermodynamical equilibrium one can 
either follow the so-called imaginary time formalism, see e.g. 
\cite{Kapusta_1989,LeBellac_1996} or the real time (Schwinger-Keldysh) 
formalism \cite{Schwinger:1960qe,Keldysh:ud}. The latter can also be 
extended to non-equilibrium situations \cite{Kad62,Bez72}. 

The perturbative expansion expressed in terms of Feynman diagrams 
allows one a systematic computation of various quantities. However, 
in order to obtain a gauge invariant finite result, one often has 
to resum a class of diagrams as required by the Hard Loop Approach 
\cite{Braaten_1990,Taylor:ia,Braaten:1991gm} (the real-time 
formulation is discussed in \cite{Carrington_1999}). The approach, 
which was first developed for equilibrium systems 
\cite{Braaten_1990,Taylor:ia,Braaten:1991gm} (for a review 
see \cite{Thoma_1995}) and then extended to the non-equilibrium 
case \cite{Pisarski:1997cp,Mrowczynski_2000,Mrowczynski:2004kv}, 
distinguishes soft from hard momenta. In the case of ultrarelativitic 
QED plasmas in equilibrium, the soft momenta are of order $eT$ while 
the hard momenta are of order $T$ with $T$ being the plasma temperature. 
One obviously assumes that $1/e \gg 1$. The Hard Loop Approach deals 
with soft collective excitations generated by hard plasma particles 
which dominate the distribution functions. 

\begin{figure}
\epsfxsize15pc 
\epsfclipon     
\centerline{\epsfbox{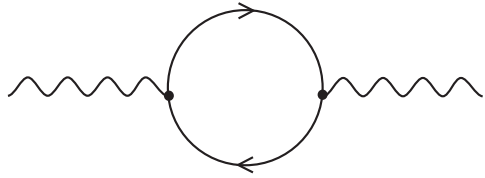}}
\caption{The lowest order contribution to the QED 
polarization tensor.}
\label{polarization}
\end{figure}

As an example, we consider the polarization tensor given by 
Equation~\ref{Pi-kinetic}, which was obtained within the 
kinetic theory in Sect.~\ref{sec-kin}. We restrict ourselves to 
ultrarelativitic QED plasmas. In the lowest order of the perturbative 
expansion, the polarization tensor or photon self-energy is given by 
the diagram shown in Figure~\ref{polarization}. The tensor can be 
decomposed into the vacuum and medium contributions. The first one 
requires an usual renormalization because of a ultraviolet divergence, 
whereas the medium part appears to be ultraviolet finite. One reproduces Equation~\ref{Pi-kinetic} applying to the diagrammatic result the Hard 
Loop Approximation which requires that the energy and momentum ($\omega$, 
${\bf k}$) of the external photon line are much smaller than the momentum 
(${\bf p}$) of the electron loop. Then, it appears that the vacuum part 
can be neglected, as it is much smaller than the medium part. In the case 
of an ultrarelativistic equilibrium EMP, Equations~\ref{polar-equi} 
were derived diagrammatically in \cite{Klimov_1982,Weldon_1982}. 
In QGP the lowest order polarization tensor (gluon self energy) 
includes one-loop diagrams with internal gluon and ghost lines. 
The final result for the gluon polarization tensor in the 
high-temperature approximation essentially coincides with the QED 
expression.  The color degrees of freedom enter through the trivial color factor $\delta_{ab}$. In the case of equilibrium
QGP, one additionally replaces in Equation~\ref{polar-equi} 
the thermal photon mass by a thermal gluon mass given by 
\be
m_g^2 = \frac {g^2T^2}{3}\> \left (1+\frac {N_f}{6}\right )\; ,
\label{gluon_mass}
\ee
where $N_f$ indicates the number of light quark flavors. 

The Hard Loop Approach can be nicely formulated in terms of an
effective action. Such an action for an equilibrium system was
derived diagrammatically in \cite{Taylor:ia} and in the 
explicitly gauge invariant form in \cite{Braaten:1991gm}. 
The equilibrium Hard Loop action was also found within the 
semiclassical kinetic theory \cite{Blaizot:1993be,Kelly:1994dh}.
The action was generalized \cite{Pisarski:1997cp,Mrowczynski:2004kv}
for non-equilibrium systems which, however, are on average locally 
color neutral, stationary and homogeneous.

The starting point was the effective action which describes an 
interaction of classical fields with currents induced by these fields 
in the plasma. The lagrangian density is quadratic in the gluon and 
quark fields and it equals
\be
\label{action-2}
{\cal L}_2(x) =  -\int d^4 y \bigg[ {1\over2} 
A^a_\mu(x) \Pi^{\mu \nu}_{ab}(x-y) A^b_\nu(y) 
+ \bar{\Psi}(x) \Sigma (x-y) \Psi (y) \bigg] \;,
\ee
where $\Pi^{\mu \nu}_{ab}$ and $\Sigma$ is the gluon polarization 
tensor and the quark self-energy, respectively, while $A^a$ and
$\Psi$ denote the gluon and quark fields. Following Braaten and
Pisarski \cite{Braaten:1991gm}, the lagrangian from 
Equation~\ref{action-2} was modified to comply with the requirement 
of gauge invariance. The final result, which is non-local but 
manifestly gauge invariant, is 
\ba
\label{HL-action}
{\cal L}_{\rm HL}(x) =  {g^2\over2} 
\int {d^3p \over (2\pi)^3} 
&\bigg[& f({\bf p)} \;
F_{\mu \nu}^a (x)
\bigg({p^\nu p^\rho \over (p \cdot D)^2} \bigg)_{ab} \;
F_\rho^{\;\;b \,\mu} (x)
\\ [2mm] \nonumber 
&+& i {N_c^2 -1 \over 4N_c}
\tilde f ({\bf p}) \;
 \bar{\Psi}(x) {p \cdot \gamma \over p\cdot D}
\Psi (x) \bigg] \;,
\ea
where $F^{\mu \nu}_a$ is the strength tensor and $D$ denotes 
the covariant derivative; $f({\bf p})$ and $\tilde f ({\bf p})$ 
are the effective parton distribution functions defined as 
$f({\bf p}) \equiv n({\bf p})+\bar n ({\bf p}) + 
2N_c n_g({\bf p})$ and $\tilde f ({\bf p}) \equiv 
n({\bf p}) + \bar n ({\bf p}) + 2 n_g({\bf p})$; 
$n({\bf p})$, $\bar n ({\bf p})$ and $n_g({\bf p})$ are the 
distribution functions of quarks, antiquarks and gluons 
of a single-color component in a homogeneous and stationary 
plasma which is locally and globally colourless; the spin 
and flavor are treated as parton internal degrees of freedom. 
The quarks and gluons are assumed to be massless. The effective 
action given by Equation~\ref{HL-action} generates $n-$point 
functions which obey the Ward-Takahashi identities. 
Equation~\ref{HL-action} holds under the assumption that the 
field amplitude is much smaller than $T/g$ where $T$ denotes 
the characteristic momentum of (hard) partons.


\section{COLLECTIVE PHENOMENA}
\label{sec-collective}
                                                                                

The most characteristic feature of the EM and QCD plasma, which
results from a long range interaction governing both systems, is 
a collective behavior which leads to specific plasma phenomena like 
screening, plasma oscillations, instabilities, etc.

Since the electromagnetic and chromodynamic polarization tensors,
which are obtained in the linear response analysis, are essentially 
the same, the collective effects in EMP and QGP are very similar 
in the linear response regime. As our discussion is limited to this 
regime, mostly the electromagnetic plasma is considered in this 
section.

\subsection{Screening}
\label{sec-screen}

We start with screening of electric charges in the plasma. To 
discuss the effect, let us consider an electric field generated by 
the point-like charge $q$ moving with a velocity  ${\bf v}$ in the 
plasma. The problem is studied in numerous plasma handbooks e.g. 
in \cite{Ichimaru_1973}. The induction vector obeys the Maxwell 
equation
$$
\nabla \cdot {\bf D}(x) = 
4\pi q \delta^{(3)}( {\bf r} - {\bf v}t ) \;,
$$
with $x \equiv (t, {\bf r})$. After the Fourier transformation, which
is defined by Equation~\ref{Fourier}, the induction vector reads
\be
\label{D1}
i {\bf k} \cdot {\bf D}(k) = 8 \pi^2 q \:
\delta(\omega - {\bf k}\cdot {\bf v}) \;,
\ee
where $k \equiv (\omega, {\bf k})$. The induction vector ${\bf D}(k)$ 
is related to the electric field ${\bf E}(k)$ through the dielectric 
tensor $\varepsilon^{ij}(k)$ as
\be
\label{D-E}
D^i(k) = \varepsilon^{ij}(k) E^j(k) \;.
\ee
We note that the dielectric tensor $\varepsilon^{ij}(k)$, which
carries information on the electromagnetic properties of a 
medium, can be expressed through the polarization tensor as
\be
\label{e-Pi}
\epsilon^{ij}(k) = \delta^{ij} + {1 \over \omega^2} \Pi^{ij}(k) \;.
\ee
In the isotropic plasma there are only two independent components 
of the dielectric tensor $\varepsilon_T $ and $\varepsilon_L$
which are related to $\varepsilon^{ij}$ as
\be
\label{e-L-T}
\varepsilon^{ij}(k) = \varepsilon_T (k )
 \Big(\delta^{ij} - k^i k^j/{\bf k}^2 \Big)
+ \varepsilon_L (k) \; k^i k^j/{\bf k}^2 \;.
\ee
Using Equations~\ref{D-E},~\ref{e-L-T}, and expressing the electric 
field ${\bf E}$ through the scalar $\phi$ and vector ${\bf A}$ 
potentials (${\bf E}(k) = -i{\bf k} \phi (k) + i \omega {\bf A}(k)$) 
in the Coulomb gauge (${\bf k} \cdot {\bf A}(k) = 0$), one finds 
the electric potential in a medium (the wake potential) 
\be 
\label{wake}
\phi (x) = 4 \pi q \int  
\frac{d^3k}{(2\pi)^3}
\frac{e^{i{\bf k}({\bf r} - {\bf v}t)}}{\varepsilon_L (\omega = {\bf v}\cdot {\bf k}, {\bf k}) \: {\bf k}^2} \;.
\ee

Let us first consider the simplest case of the potential generated 
by a static (${\bf v} = 0$) charge. Using Equations~\ref{pol-L-T}, 
\ref{polar-equi}, $\varepsilon_L (0, {\bf k})$ of an
ultrarelatvistic electron-positron plasma is found as
\be
\varepsilon_L (0, {\bf k}) = 
1 + \frac{m_D^2}{{\bf k}^2} \;,
\ee
where $m_D$ is the so-called Debye mass given by 
$m^2_D = e^2T^2/3 = 3m_\gamma^2 = \Pi_L(0,{\bf k})$. Then,
Equation~\ref{wake} gives the well-known screened potential
\be
\phi ({\bf r}) = \frac{q}{r} \, e^{-m_D r} \;,
\ee
with $r \equiv |{\bf r}|$. As seen, the inverse Debye mass has the 
interpretation of the screening length of the potential. Since the 
average inter-particle spacing in the ultrarelatvistic plasma is 
of order $T^{-1}$, the number of particles in the Debye sphere 
(the sphere of the radius $m_D^{-1}$) is of order $e^{-3}$ which 
is, as already mentioned in the Introduction, much larger that 
than unity in the weakly coupled plasma ($1/e^2 \gg 1$). This
explains the collective behavior of the plasma, as motion of
particles from the Debye sphere is highly correlated.

For ${\bf v} \not= 0$ the potential given by Equation~\ref{wake} has 
a rich structure. In the context of QGP it has been discussed in 
\cite{Mustafa_2005a,Ruppert_2005,Mustafa_2006}, showing that it 
can exhibit attractive contributions even between like-sign charges 
in certain directions \cite{Mustafa_2005a}. For a supersonic particle, 
the potential can reveal a Mach cone structure associated with 
\u Cerenkov radiation when electromagnetic properties of the plasma 
are appropriately modeled \cite{Ruppert_2005,Mustafa_2006}.

\subsection{Collective modes}
\label{sec-dis-eq}

Let us consider a plasma in a homogenous, stationary state with 
no local charges, no currents. As a fluctuation or perturbation of 
this state, there appear local charges or currents generating electric 
and magnetic fields which in turn interact with charged plasma 
particles. Then, the plasma reveals a collective motion which 
classically is termed as plasma oscillations. Quantum-mechanically
we deal we quasi-particle collective excitations of the plasma.

The collective modes are found as solutions of the dispersion
equation obtained from the equation of motion of the Fourier 
transformed electromagnetic potential $A^{\mu}(k)$ which is
\be
\label{eq-A}
\Big[ k^2 g^{\mu \nu} -k^{\mu} k^{\nu} - \Pi^{\mu \nu}(k) \Big]
A_{\nu}(k) = 0 \;,
\ee
where the polarization tensor $\Pi^{\mu \nu}$ contains all dynamical 
information about the system. The general dispersion equation is then
\be
\label{dispersion-pi}
{\rm det}\Big[ k^2 g^{\mu \nu} -k^{\mu} k^{\nu} - \Pi^{\mu \nu}(k) \Big]
 = 0 \;.
\ee

Due to the transversality of $\Pi^{\mu \nu}(k)$ not all components 
of $\Pi^{\mu \nu}(k)$ are independent from each other, and
consequently the dispersion Equation~\ref{dispersion-pi}, which
involves a determinant of a $4\times4$ matrix, can be simplified 
to the determinant of a $3\times3$ matrix. For this purpose one
usually introduces the dielectric tensor $\varepsilon^{ij}(k)$
related to the polarization tensor by Equation~\ref{e-Pi}.
Then, the dispersion equation gets the form
\be
\label{dispersion-g}
{\rm det}\Big[ {\bf k}^2 \delta^{ij} -k^i  k^j
- \omega^2 \varepsilon^{ij}(k)  \Big]  = 0 \,.
\ee
The relationship between Equation~\ref{dispersion-pi} and
Equation~\ref{dispersion-g} is most easily seen in the Coulomb
gauge when $\phi = 0$ and ${\bf k} \cdot {\bf A}(k)=0$. Then,
${\bf E} = i\omega {\bf A}$ and Equation~\ref{eq-A} is immediately
transformed into an equation of motion of ${\bf E}(k)$ which
further provides the dispersion Equation~\ref{dispersion-g}.

As expressed by Equation~\ref{e-L-T}, there are only two independent 
components of the dielectric tensor ($\varepsilon_T (k)$ and 
$\varepsilon_L (k)$) in the isotropic plasma. Then, the dispersion 
Equation~\ref{dispersion-g} splits into  two equations
\be
\label{dis-eq-iso}
\varepsilon_T (k) = {\bf k}^2/ \omega^2 \;,\;\;\;\;\;\;\;\;\;\;\;
\varepsilon_L (k) = 0 \;.
\ee

Solutions of the dispersion equations $\omega ({\bf k})$, with
a complex, in general, frequency $\omega$, represent plasma modes
which classically are, as already mentioned, the waves of electric 
and/or magnetic fields in the plasma while quantum-mechanically 
the modes are quasi-particle excitations of the plasma system. If the 
imaginary part of the mode's frequency $\Im \omega$ is negative, 
the mode is damped - its amplitude exponentially decays in time as 
$e^{\Im \omega t}$. When $\Im \omega =0$ we have a stable mode 
with a constant amplitude. Finally, if $\Im \omega > 0$, the mode's 
amplitude exponentially grows in time - there is an instability.

When the electric field of a mode is parallel to its wave vector
${\bf k}$, the mode is called {\it longitudinal}. A mode is called
{\it transverse} when the electric field is transverse to the
wave vector. The Maxwell equations show that the longitudinal modes,
also known as {\it electric}, are associated with electric 
charge oscillations; the transverse modes, also known as 
{\it magnetic}, are associated with electric current oscillations.

\begin{figure}
\epsfxsize20pc         
\epsfclipon     
\centerline{\epsfbox{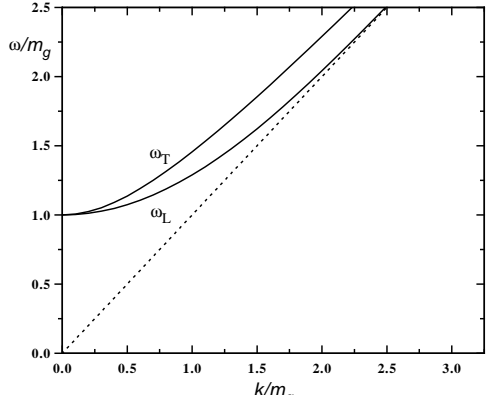}}
\caption{Dispersion relation of longitudinal and transverse plasma waves.}
\label{fig-dispersion}
\end{figure}

The collective (boson) modes in the equilibrium ultrarelativistic plasma 
are shown in Figure~\ref{fig-dispersion}. There are longitudinal modes 
also called plasmons and transverse modes. Both start at zero momentum at 
the plasma frequency which is identical to the thermal photon (or gluon) 
mass, $\omega_p=m_\gamma$. The dispersion relations lies above the light 
cone ($\omega >|{\bf k}|$) showing that the plasma waves are undamped 
(no Landau damping) in the high-temperature limit. As explained in 
Sect.~\ref{sec-elec-mecha}, the Landau damping, which formally arises 
from the imaginary part of the polarization tensor given by 
Equation~\ref{polar-equi} at $\omega^2 < {\bf k}^2$, occurs when the 
energy of the wave is transferred to plasma particles moving with 
velocity equal to the phase velocity ($\omega/|{\bf k}|$) of the wave. 
If the phase velocity is larger than the speed of light such a transfer 
is obviously not possible.

\subsubsection{TWO-STREAM SYSTEM}
\label{2-streams}

As an example of the rich spectrum of collective modes, we consider 
the two-stream system within the hydrodynamic approach when the effect 
of pressure gradients is neglected. Details of the analysis can be found
in \cite{Manuel:2006hg}. The dielectric tensor provided by the polarization 
tensor from Equation~\ref{Pi2} is
\be
\label{epsilon-hydro}
\varepsilon^{ij}(\omega,{\bf k}) =
\Big(1 - \frac{\omega_p^2}{\omega^2} \Big) \delta^{ij}
- \frac{4\pi}{\omega^2}
\sum_\alpha
\frac{q_\alpha^2 \bar n_\alpha^2}{\bar \epsilon_\alpha + \bar p_\alpha}
\bigg[ \frac{\bar v_\alpha^i k^j +\bar v_\alpha^j k^i}
{\omega - {\bf k} \cdot \bar {\bf v}_\alpha}
- \frac{(\omega^2 - {\bf k}^2) \bar v_\alpha^i \bar v_\alpha^j}
{(\omega - {\bf k} \cdot \bar {\bf v}_\alpha )^2} \bigg] \;,
\ee
where $\bar {\bf v}_\alpha$ is the hydrodynamic velocity related
to the hydrodynamic four-velocity $\bar u^\mu_\alpha$; $\omega_p$ 
is the plasma frequency given as
\be
\omega_p^2 \equiv 4\pi \sum_\alpha
\frac{ q_\alpha^2\bar n_\alpha^2}{\bar \epsilon_\alpha + \bar p_\alpha} \;.
\ee

The index $\alpha$, which labels the streams and plasma components, has 
four values, $\alpha = L-, L+, R-,R+$. The first character labels the 
stream (`$R$' for right and `$L$' for left) while the second one labels
the plasma component (`$+$' for positive and `$-$' for negative). For 
simplicity we assume here that the streams are neutral and identical to 
each other and their velocities, which are chosen along the axis $z$, 
are opposite to each other. Then, 
\ba
\nonumber 
\bar n &\equiv& \bar n_{L-} = \bar n_{L+} = \bar n_{R-} = \bar n_{R+}  \;, 
\;\;\;\;\;\;\;
\bar \epsilon \equiv \bar \epsilon_{L-} = \bar \epsilon_{L+} 
= \bar \epsilon_{R-} = \bar \epsilon_{R+} \;, 
\\ \nonumber 
\bar p &\equiv& \bar p_{L-} = \bar p_{L+} = \bar p_{R-} = \bar p_{R+}\;,
\;\;\;\;\;\;\;
\bar v \equiv \bar v_{L-} = \bar v_{L+} = -\bar v_{R-} = -\bar v_{R+}\;, 
\\ \label{streams}
e &=& q_{L-} = - q_{L+} = q_{R-} = -q_{R+} \;,
\ea
and the plasma frequency equals 
$\omega_p^2 = 16\pi e^2 \bar n^2 /(\bar \epsilon + \bar p)$. 

The wave vector is first chosen to be parallel to the axis $x$,
${\bf k} = (k,0,0)$. Due to Equations~\ref{streams}, the off-diagonal 
elements of the matrix in Equation~\ref{dispersion-g} vanish and the 
dispersion equation with the dielectric tensor given by 
Equation~\ref{epsilon-hydro} is
\be
(\omega^2 - \omega_p^2)(\omega^2 - \omega_p^2 - k^2)
\Big(\omega^2 - \omega_p^2 - k^2
- \lambda^2  \; \frac{k^2 - \omega^2}{\omega^2} \Big) = 0 \;,
\ee
where $\lambda^2 \equiv \omega_p^2 \bar v^2$. As solutions of the
equation, one finds a stable longitudinal mode with
$\omega^2 = \omega_p^2$ and a stable transverse mode with
$\omega^2 = \omega_p^2 + k^2$. There are also transverse modes with
\be
\label{solut11}
\omega_{\pm}^2 = \frac{1}{2}\Big[
\omega_p^2 - \lambda^2 + k^2 \pm
\sqrt{(\omega_p^2 - \lambda^2 + k^2)^2 + 4 \lambda^2 k^2} \, \Big]
\;.
\ee
As seen, $\omega_+^2 > 0$ but $\omega_-^2 < 0$. Thus, the mode $\omega_+$
is stable and there are two modes with pure imaginary frequency corresponding
to $\omega_-^2 < 0$. The first mode is overdamped while the second one
is the well-known unstable Weibel mode \cite{Wei59} leading to the 
filamentation instability. A physical mechanism of the instability 
is explained in Sect.~\ref{sec-mag-mecha}.

The wave vector, as the stream velocities, is now chosen along the 
$z-$axis {\it i.e.} ${\bf k} = (0,0,k)$. Then, the matrix in 
Equation~\ref{dispersion-g} is diagonal. With the dielectric tensor 
given by Equation~\ref{epsilon-hydro}, the dispersion equation reads
\be
\label{disper-eq22}
(\omega^2 - \omega_p^2 - k^2)^2\bigg\{
\omega^2 - \omega_p^2
- \omega_p^2 \Big[\frac{ k \bar v} {\omega - k \bar v} +
\frac{(k^2 - \omega^2)\bar v^2} {2(\omega - k \bar v)^2}
- \frac{ k\bar v}{\omega + k \bar v} +
\frac{(k^2 - \omega^2)\bar v^2} {2(\omega + k \bar v )^2} \Big]
\bigg\}= 0 \;.
\ee
There are two transverse stable modes with $\omega^2 = \omega_p^2 + k^2$.
The longitudinal modes are solutions of the above equation which can be
rewritten as
\be
\label{disper-2stream}
1 - \omega_0^2\bigg[
\frac{1}{(\omega - k \bar v)^2}
+\frac{1}{(\omega + k \bar v )^2} \bigg] = 0 \;,
\ee
where $\omega_0^2 \equiv \omega_p^2/2\bar \gamma^2$ with 
$\bar \gamma = (1-{\bar v}^2)^{-1/2}$.
With the dimensionless quantities $x\equiv \omega/\omega_0$,
$y \equiv k \bar v/ \omega_0$, Equation~\ref{disper-2stream} is
\be
(x^2 - y^2)^2 - 2x^2 - 2y^2 = 0 \;,
\ee
and it is solved by
\be
x_\pm^2 = y^2 + 1 \pm \sqrt{4y^2 + 1} \;.
\ee
As seen, $x_+^2$ is always positive and thus, it gives two
real (stable) modes; $x_-^2$ is negative for $ 0 < y < \sqrt{2}$
and then, there are two pure imaginary modes. The unstable one
corresponds to the two-stream electrostatic instability.
A physical mechanism of the instability is explained in 
Sect.~\ref{sec-elec-mecha}.

\subsection{Instabilities}

Presence of unstable modes in a plasma system crucially influences
its dynamics. Huge difficulties encountered by the half-a-century
program to built a thermonuclear reactor are just related to various
instabilities experienced by a plasma which make the system's
behavior very turbulent, hard to predict and hard to control. 

\begin{figure}
\epsfxsize25pc
\epsfclipon     
\centerline{\epsfbox{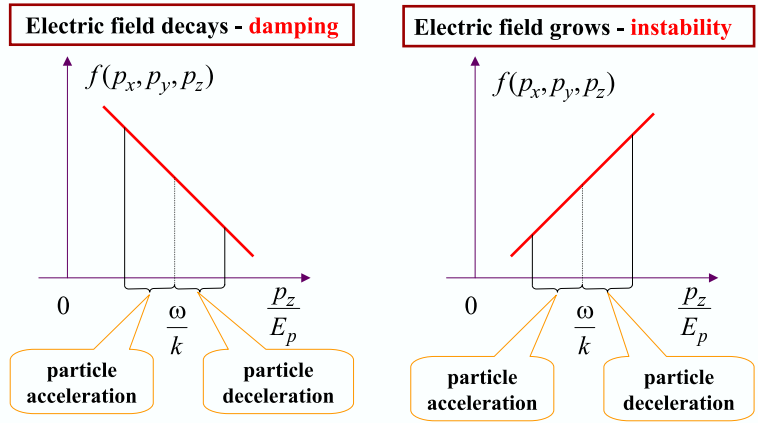}}
\caption{The mechanism of energy transfer between particles
and fields.}
\label{fig-elec-mechanism}
\end{figure}

There is a large variety of instabilities - the history of plasma
physics is said to be a history of discoveries of new and new
instabilities. Plasma instabilities can be divided into two general
groups (1) {\it hydrodynamic} instabilities, caused by coordinate
space inhomogeneities, (2) {\it kinetic} instabilities due to 
non-equilibrium momentum distribution of plasma particles. 

The hydrodynamic instabilities are usually associated with
phenomena occurring at the plasma boundaries. In the case of
QGP, this is the domain of highly non-perturbative QCD where
nonAbelian nature of the theory is of crucial importance.
Then, the behavior of QGP is presumably very different
from that of EMP, and thus, we will not speculate about
possible analogies.  

The kinetic instabilities are simply the collective modes with 
positive $\Im \omega$ introduced in Sect.~\ref{sec-dis-eq} and found 
in Sect.~\ref{2-streams} in the specific case of the two-stream 
system. Thus, we have longitudinal (electric) and transverse 
(magnetic) instabilities. In the non-relativistic plasma the 
electric instabilities are usually much more important than the
magnetic ones, as the magnetic effects are suppressed by the factor
$v^2/c^2$ where $v$ is the particle's velocity. In the relativistic
plasma both types of instabilities are of similar strength. As will
be discussed later on, the electric instabilities occur when the
momentum distribution of plasma particles has more than one maximum,
as in the two-stream system. A sufficient condition for the magnetic
instabilities appears to be an anisotropy of the momentum
distribution.

\subsubsection{MECHANISM OF ELECTRIC INSTABILITY}
\label{sec-elec-mecha}

Let us consider a plane wave of the electric field with the wave vector 
along the $z$ axis. For a charged particle, which moves with a velocity 
$v= p_z/E_p$ equal to the phase velocity of the wave $v_\phi= \omega/k$, 
the electric field does not oscillates but it is constant. The particle 
is then either accelerated or decelerated depending on the field's 
phase. For an electron with $v=v_\phi$ chances to be accelerated and 
to be decelerated are equal to each other, as the time intervals spent 
by the particle in the acceleration zone and in the deceleration 
zone are equal to each other.

\begin{figure}
\epsfxsize20pc         
\epsfclipon\centerline{\epsfbox{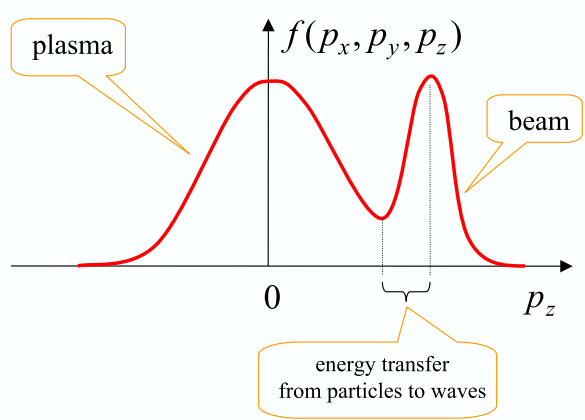}}
\caption{Momentum distribution of the plasma-beam system.}
\label{fig-plasma-beam}
\end{figure}

Let us now consider electrons with the velocities somewhat smaller
the phase velocity of the wave. Such particles spend more time in the 
acceleration zone than in the deceleration zone, and the net result 
is that the particles with $v < v_\phi$ are accelerated. Consequently, 
the energy is transferred from the electric field to the particles. 
The particles with  $v > v_\phi$ spend more time in the deceleration 
zone than in the acceleration zone, and thus they are effectively 
decelerated - the energy is transferred from the particles to the field. 
If the momentum distribution is such that there are more electrons
in a system with  $v < v_\phi$ than with $v > v_\phi$, the wave
looses energy which is gained by the particles, as shown in the 
left-hand-side of Figure~\ref{fig-elec-mechanism}. This is the mechanism 
of famous collisionless Landau damping of the plasma oscillations. If 
there are more particles with $v > v_\phi$ than with $v < v_\phi$, 
the particles loose energy which is gained by the wave, as in the 
right-hand-side of Figure~\ref{fig-elec-mechanism}. Consequently, 
the wave amplitude grows. This is the mechanism of electric instability. 
As explained above, it requires the existence of the momentum interval 
where $f_n({\bf p})$ grows with ${\bf p}$. Such an interval appears 
when the momentum distribution has more than one maximum. This happens
in the two-stream system discussed in Sect.~\ref{2-streams} or in the
system of a plasma and a beam shown in Figure~\ref{fig-plasma-beam}.  

\subsubsection{MECHANISM OF MAGNETIC INSTABILITY}
\label{sec-mag-mecha}
 
Since the magnetic instabilities appear to be relevant for QGP 
produced in relativistic heavy-ion collisions (see below), we 
discuss them in more detail. Let us first explain following 
\cite{Mrowczynski:1996vh}, how the unstable transverse modes are 
initiated. For this purpose we consider a plasma system which is 
homogeneous but the momentum distribution of particles is not of 
the equilibrium form, it is {\em not} isotropic. The system is 
on average locally neutral ($\langle j^{\mu} (x)\rangle = 0$) but 
current fluctuations are possible, and thus the correlator 
$\langle j^{\mu} (x_1) j^{\nu}(x_2) \rangle$
is in general non-zero. Since the plasma is assumed to be weakly
coupled, the correlator can be estimated neglecting the interaction
entirely. Then, when the effects of quantum statistics are also
neglected, the correlator equals
\ba
\label{cur-cor-x}
M^{\mu \nu} (t,{\bf x}) \buildrel \rm def \over =
\langle j^{\mu}(t_1,{\bf x}_1) j^{\nu} (t_2,{\bf x}_2) \rangle
= \sum_n q_n^2 \int \frac{d^3p}{(2\pi)^3}\;
{p^{\mu} p^{\nu} \over E_p^2} \;
f_n({\bf p}) \; \delta^{(3)} ({\bf x} -{\bf v} t)  \;,
\ea
where ${\bf v} \equiv {\bf p}/ E_p$ and 
$(t,{\bf x}) \equiv (t_2-t_1,{\bf x}_2-{\bf x}_1)$. Due to the average 
space-time homogeneity, the correlator given by Equation~\ref{cur-cor-x} 
depends only on the difference $(t_2-t_1,{\bf x}_2-{\bf x}_1)$. The 
space-time points $(t_1,{\bf x}_1)$ and $(t_2,{\bf x}_2)$ are correlated 
in the system of non-interacting particles if a particle travels from 
$(t_1,{\bf x}_1)$ to $(t_2,{\bf x}_2)$. For this reason the delta  
$\delta^{(3)} ({\bf x} - {\bf v} t)$ is present in 
Equation~\ref{cur-cor-x}. The sum and momentum integral represent 
summation over all particles in the system. The fluctuation spectrum 
is found as the Fourier transform of Equation~\ref{cur-cor-x} 
{\it i.e.}
\be
\label{cur-cor-k}
M^{\mu \nu} (\omega ,{\bf k}) = 
\sum_n q_n^2 \int \frac{d^3p}{(2\pi)^3}\;
{p^{\mu} p^{\nu} \over E_p^2} \;
f_n({\bf p}) \; 2\pi \delta (\omega -{\bf kv}) \;.
\ee

\begin{figure}
\epsfxsize20pc
\epsfclipon              
\centerline{\epsfbox{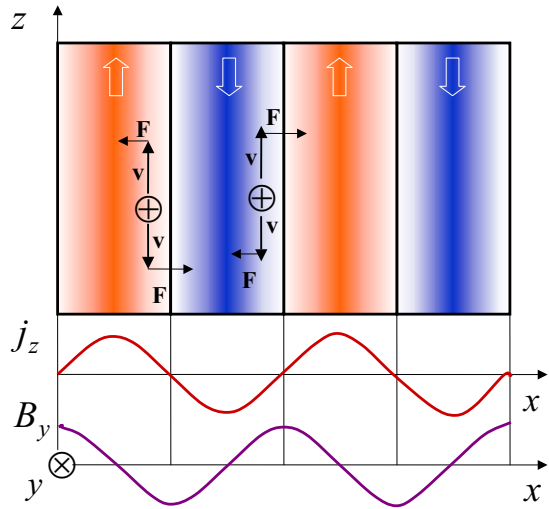}}
\caption{The mechanism of filamentation instability, 
see text for a description.}
\label{fig-mag-mechanism}
\end{figure}

To further study the fluctuation spectrum, the particle's momentum
distribution has to be specified. We will present here only a
qualitative discussion of  Equations~\ref{cur-cor-x},~\ref{cur-cor-k},
assuming that the momentum distribution is strongly elongated in one 
direction which is chosen to be along the $z$ axis. Then, the correlator 
$M^{zz}$ is larger than $M^{xx}$ or $M^{yy}$. It is also clear that
$M^{zz}$ is the largest when the wave vector ${\bf k}$ is along 
the direction of the momentum deficit, as in such a case the delta
function $\delta (\omega -{\bf kv})$ does not much constrain the
integral in Equation~\ref{cur-cor-k}. Since the momentum distribution 
is elongated in the $z$ direction, the current fluctuations are 
the largest when the wave vector ${\bf k}$ is the $x\!-\!y$ plane.
Thus, we conclude that some fluctuations in the anisotropic system
are large, much larger than in the isotropic one. An anisotropic
system has a natural tendency to split into the current filaments
parallel to the direction of the momentum surplus. These currents 
are seeds of the transverse unstable mode known as the filamentation
or Weibel instability \cite{Wei59} which was found in the two-stream
system discussed in Sect.~\ref{2-streams}.

Let us now explain in terms of elementary physics why the fluctuating
currents, which flow in the direction of the momentum surplus, can
grow in time. The form of the fluctuating current is chosen to be
\be
\label{flu-cur}{\bf j}(x) = j \: \hat {\bf e}_z \: {\rm cos}(k_x x) \;,
\ee
where $\hat {\bf e}_z$ is the unit vector in the $z$ direction.
As seen in Equation~\ref{flu-cur}, there are current filaments of
the thickness $\pi /\vert k_x\vert$ with the current flowing in the
opposite directions in the neighboring filaments. The magnetic field 
generated by the current from Equation~\ref{flu-cur} is given as
\ban
{\bf B}(x) = 4\pi \:{j \over k_x} \: 
\hat {\bf e}_y \: {\rm sin}(k_x x) \;,
\ean
and the Lorentz force acting on the particles, which fly along 
the $z$ direction, equals
\ban
{\bf F}(x) = q \: {\bf v} \times {\bf B}(x) =
- 4\pi q \: v_z \: {j \over k_x} \: 
\hat {\bf e}_x \: {\rm sin}(k_x x) \;,
\ean
where $q$ is the particle's electric charge. One observes, see 
Figure~\ref{fig-mag-mechanism}, that the force distributes the particles 
in such a way that those, which positively contribute to the current 
in a given filament, are focused in the filament center while those,
which  negatively contribute, are moved to the neighboring one. 
Thus, the initial current is growing and the magnetic field generated
by this current is growing as well. The instability is driven by the
the energy transferred from the particles to fields. More specifically, 
the kinetic energy related to the motion along the direction of the 
momentum surplus is used to generate the magnetic field.

\subsubsection{ROLE OF INSTABILITIES}
\label{sec-role-insta}

As already mentioned, there is a large variety of plasma instabilities 
which strongly influence numerous plasma characteristics. Not much is 
known of the hydrodynamic instabilities of QGP and if they exist, they 
belong to the highly non-perturbative sector of QCD which is still 
poorly understood.  As explained in Sect.~\ref{sec-elec-mecha}, the 
electric instabilities occur in a two-stream system, or more generally, 
in systems with a momentum distribution of more than one maximum. While 
such a distribution is common in EMP, it is rather irrelevant for QGP 
produced in relativistic heavy-ion collisions where the global as well 
as local momentum distribution is expected to monotonously decrease 
in every direction from the maximum. The electric instabilities are 
absent in such a system but a magnetic unstable mode, which has been
discussed in Sect.~\ref{sec-mag-mecha}, is possible. The filamentaion 
instability was first argued to be relevant for the QGP produced in 
relativistic heavy-ion collisions in 
\cite{Mrowczynski:1993qm,Mrowczynski:xv,Mrowczynski:1996vh}. A  
characteristic time of instability growth was estimated \cite{Mrowczynski:1993qm,Mrowczynski:xv} to be shorter or at least 
comparable to other time scales of the parton system evolution. 
The mechanism of instability growth was also clarified 
\cite{Mrowczynski:1996vh}. The early arguments were substantiated in 
the forthcoming analytic calculations 
\cite{Randrup:2003cw,Romatschke:2003ms,Arnold:2003rq} 
and numerical simulations 
\cite{Rebhan:2004ur,Arnold:2004ih,Dumitru:2005gp,Arnold:2005vb,Rebhan:2005re}.

A main consequence of instabilities is a fast equilibration of
the weakly coupled plasma. The problem is of particular interest
as the experimental data on heavy-ion collisions, where QGP
production is expected, suggest that an equilibration time 
of the parton system is below 1 ${\rm fm}/c$ \cite{Heinz:2004pj}.
A whole scenario of instabilities driven equilibration is reviewed 
in \cite{Mrowczynski:2005ki}. Here we only mention the main points, 
starting with an observation that collisions of charged particles 
are not very efficient in redistributing particle momenta, as the 
Rutherford cross section is strongly peaked at a small momentum 
transfer. One either needs many frequent but soft collisions
or a few rare but hard collisions to substantially change a particle's
momentum. As a result the inverse time of collisional equilibration 
of QGP is of order $g^4 {\rm ln}(1/g)\,T$ \cite{Arnold:1998cy} 
where $T$ is the characteristic momentum of quarks or gluons. It 
appears that the momentum distribution is isotropized due to 
instabilities within the inverse time of order $gT$ \cite{Arnold:2004ti}. 
If $1/g \gg 1$, the collisional equilibration is obviously much slower. As discussed in Chapter~\ref{sec-strong}, the situation changes in 
strongly-coupled plasmas.

When the instabilities grow, the system becomes more and more 
isotropic \cite{Mrowczynski:xv,Arnold:2004ti} because the Lorentz 
force changes the particle's momenta and the growing fields carry an 
extra momentum. To explain the mechanism let us assume that initially 
there is a momentum surplus in the $z$ direction. The fluctuating current 
flows in the $z$ direction with the wave vector pointing in the $x$ 
direction. Since the magnetic field has a $y$ component, the Lorentz 
force, which acts on partons flying along the $z$ axis, pushes the partons 
in the $x$ direction where there is a momentum deficit. The numerical 
simulation \cite{Dumitru:2005gp} shows that growth of the instabilities 
is indeed accompanied with the system's fast isotropization. 

The system isotropizes not only due to the effect of the Lorentz
force but also due to the momentum carried by the growing field. 
When the magnetic and electric fields are oriented along the
$y$ and $z$ axes, respectively, the Poynting vector points in
the direction $x$ that is along the wave vector. Thus, the momentum 
carried by the fields is oriented in the direction of the momentum 
deficit of particles.

Although the scenario of instabilities driven equilibration looks
very promising, the problem of thermalization of QGP produced in 
heavy-ion collisions is far from being settled. It has been shown 
\cite{Schenke_2006} that inter-parton collisions, which have been 
modeled using the BGK collision term \cite{Alexandrov_1984}, reduce 
the growth of instabilities and thus slow down the process of 
equilibration. The equilibration is also slowed down due to 
expansion of QGP into vacuum 
\cite{Romatschke:2005pm,Romatschke:2006wg} which is 
a characteristic feature of QGP produced in relativistic 
heavy-ion collisions. Finally, the late stage of instability 
development, when nonAbelian effects are crucially important, 
appears to be very complex \cite{Arnold:2005qs,Dumitru:2006pz} 
and it is far from being understood. 

As already mentioned, instabilities influence various plasma 
characteristics. In particular, it is known \cite{Abe_1980} 
that turbulent magnetic fields generated in the systems, which are 
unstable with respect to transverse modes, are responsible for 
a reduction of plasma viscosity. Then, an anomalously small viscosity,
which is usually associated with strongly coupled systems, can 
occur in weakly coupled plasmas as well. Recently, it has been 
argued \cite{Asakawa:2006tc,Asakawa:2006jn} that the mechanism 
of viscosity reduction is operative in the unstable QGP.

\subsection{Energy loss}

A charged particle which moves across the plasma changes its energy 
due to several processes \cite{Ichimaru_1973}. When the particle's 
energy ($E$) is comparable to the plasma temperature ($T$), the particle 
can gain energy due to interactions with field fluctuations. (In context 
of QGP the problem was studied in \cite{Chakraborty:2006db}.) A fast
particle with $E \gg T$ looses its energy and main contributions come 
from collisions with other plasma particles and from radiation. 
In the following we discuss the energy loss of a fast particle, as 
the problem is closely related to jet quenching which was suggested 
long ago as a signature for the QGP formation in relativistic heavy-ion 
collisions \cite{Bjorken_1982,Gyulassy_1990}. 

Let us start with the collisional energy loss. The particle's collisions 
are split into two classes: {\it hard} with high-momentum transfer, 
corresponding to the collisions with plasma particles and, {\it soft} 
with low-momentum transfer dominated by the interactions with plasma 
collective modes. The momentum is called `soft' when it is of order 
of the Debye mass, $m_D$, or smaller and it is `hard' when it is larger 
than $m_D$.

The soft contribution to the energy loss, which can be treated in 
a classical way, is often called the process of plasma polarization. 
It leads to the energy loss per unit time given by the formula
\be
\label{eloss-p1}
{\Bigg( {dE \over dt} \Bigg)}_{\rm soft} 
= \int d^3x \; {\bf j}(x) {\bf E} (x) \;,
\ee
where ${\bf E}$ is the electric field induced in the plasma by the 
particle's current ${\bf j}$ which is of the form 
${\bf j}(x) = q {\bf v} \delta ^{(3)}({\bf x} - {\bf v}t)$. 
The field can be calculated by means of the Maxwell equations. After
eliminating the magnetic field, one finds the equation
$$
\Bigg[ \varepsilon^{ij}(k) - {{\bf k}^2 \over \omega ^2}
\Bigg( \delta^{ij} - {k^i k^j \over {\bf k}^2 } \Bigg) \Bigg]
E_j(k) = {4\pi \over i\omega} \; j^i(k) \;. 
$$
Since we consider equilibrium plasma, which is isotropic, 
one introduces the longitudinal ($\varepsilon_L$) and transverse 
($\varepsilon_L$) components of $\varepsilon^{ij}$. Then, 
Equation~\ref{eloss-p1} is manipulated to
\ba
\label{eloss-p2}
{\Bigg( {dE \over dx} \Bigg)}_{\rm soft}
= - {4\pi ie^2 \over v} \int {d^3k \over (2\pi )^3}
\bigg\{{ \omega \over {\bf k}^2 \varepsilon _L(k)}
+ {{\bf v}^2 - \omega ^2 /{\bf k}^2 \over 
\omega [\epsilon _T(k) -{\bf k}^2 /\omega ^2]} \bigg\}  \;,
\ea
which gives the energy loss per unit length. This formula describes
the effect of medium polarization. However, three comments are in 
order here. 

\begin{enumerate}

\item
Equation~\ref{eloss-p2} includes the charge self-interaction signaled 
by the ultraviolet divergence of the integral from Equation~\ref{eloss-p2}. 
The self-interaction is removed by subtracting from Equation~\ref{eloss-p2} 
the vacuum expression with $\varepsilon_L = \varepsilon_T = 1$. 

\item
Poles of the function under the integral Equation~\ref{eloss-p2} 
correspond to the plasma collective modes as given by the dispersion 
Equations~\ref{dis-eq-iso}. Therefore, the explicit expressions of 
$\varepsilon_L$ and $\varepsilon_L$ are not actually needed to 
compute the integral in Equation~\ref{eloss-p2}. The knowledge of 
the spectrum of quasiparticles appears to be sufficient.

\item
Equation~\ref{eloss-p2} is derived in the classical approximation
which breaks down for a sufficiently large ${\bf k}$. Therefore,
an upper cut-off is needed. The interaction with ${\bf k}$ above 
the cut-off, which, as already mentioned, is of order of the Debye mass, 
should be treated as hard collisions with plasma particles.

\end{enumerate}

The energy loss per unit length due to hard collisions is\be
\label{eloss-c}
{\Bigg( {dE \over dx} \Bigg)}_{\rm hard}  = \sum _i 
\int {d^3 k \over (2\pi )^3 } \; n_i(k) \; [{\rm flux \; factor}] 
\int d\Omega {d \sigma ^i \over d\Omega} \nu  \;,
\ee
where the sum runs over particle species distributed according 
to $n_i(k)$, $\nu \equiv E - E'$ is the energy transfer, and 
$d \sigma ^i / d\Omega$ is the respective differential cross section.

Combining Equations \ref{eloss-p2} and \ref{eloss-c}, one finds
the complete collisional energy loss. The calculations of the energy 
loss of a fast parton in the QGP along the lines presented above were 
performed in \cite{Thoma_1991,Mrowczynski_1991}. Systematic calculations 
of the collisional energy loss using the Hard Thermal Loop resummation 
technique were given in \cite{Braaten_1991a,Braaten_1991b} with a result 
which is infrared finite, gauge invariant, and complete to leading order.
Recently, the calculations of the collisional energy loss have been 
extended to anisotropic QGP \cite{Romatschke:2004au}.

It was realized that a sizeable contribution to the quark's 
energy loss comes from radiative processes \cite{Gyulassy:1993hr}. 
The problem, however, appeared to be very complex because quark's 
successive interactions in the plasma cannot be treated as independent 
from each other and there is a destructive interference of radiated 
gluons known as the Landau-Pomeranchuk-Migdal effect \cite{LPM}. There 
are numerous papers devoted to the radiative energy loss and the whole 
problem is reviewed in \cite{Baier_2000}. A general conclusion of these
studies is that the energy lost by a fast light quark quadraticaly 
(not linearly) depends on the path traversed in QGP, as the radiative 
energy loss dominates over the collisional. Recent experiments at RHIC 
show \cite{Adler_2006,Abelev_2006}, however, that heavy quarks, whose
radiative energy loss is significantly suppressed, are strongly dragged  
in the QGP medium. It may suggest that the collisional energy loss
should be actually enhanced as theoretically predicted in 
\cite{Mustafa_2005b}.

\section{STRONGLY COUPLED PLASMAS}
\label{sec-strong}

Our discussion of the collective phenomena presented in
Sect.~\ref{sec-collective} was limited to the weakly interacting plasmas
with the coupling constant much smaller than unity. However, QGP produced 
in ultrarelativistic heavy-ion collisions is presumably strongly coupled 
(sQGP) as the temperature is never much larger than $\Lambda_{\rm QCD}$, 
and the regime of asymptotic freedom is not reached. QGP is certainly a 
strongly interacting system close to the confinement phase transition.
There are indeed hints in the extensive experimental material collected
at RHIC \cite{Arsene:2004fa,Back:2004je,Adams:2005dq,Adcox_2005}
that the matter produced at the early stage of nucleus-nucleus 
collisions RHIC is in the form of sQGP for a few fm/c. In particular, 
the characteristics of elliptic flow and particle spectra, which are 
described well by ideal hydrodynamics, seem to indicate a fast 
thermalization and small viscosity of the plasma. Both features 
are naturally explained assuming a strong coupling of the plasma 
\cite{Heinz:2004pj,Gyulassy_2005,Shuryak:2005pp,Cassing_2005,Thoma_2006}. 

Although a fast thermalization \cite{Mrowczynski:2005ki} as well as 
a small viscosity \cite{Asakawa:2006tc,Asakawa:2006jn} can also be 
explained by instabilities, the idea of sQGP certainly needs to be 
examined. However, the theoretical tools presented in 
Chapter~\ref{sec-tools} implicitly or explicitly assume small coupling 
constant and they are of limited applicability. A powerful approach, 
which can be used to study sQGP is the lattice formulation of QCD, 
for a review see \cite{Karsch:2001cy}. However, lattice QCD calculations, 
which are mostly numerical, encounter serious problems to incorporate 
quark degrees of freedom. It is also very difficult to analyze time 
dependent plasma characteristics. 

Strongly coupled conformal field theories such as supersymmetric QCD
can be studied by means of the so-called AdS/CFT duality 
\cite{Maldacena:1997re}. Although some very interesting results 
on the conformal QGP were obtained in this way, see e.g. 
\cite{Kovtun:2004de} and references therein, relevance of these
results for QGP governed by QCD, not by supersymmetric QCD, is 
unclear. Thus, the question arises what we can learn about sQGP 
from strongly coupled EMP.

We first note that most of EMP in nature and technological applications 
are weakly coupled i.e. the interaction energy between the plasma particles 
is much smaller than their thermal (kinetic) energy. This is because strongly
coupled plasmas require a high particle density and/or low temperature, 
at which usually strong recombination occurs and the plasma state vanishes. 
Exceptions are the ion component in white dwarfs, metallic hydrogen and
other states of dense warm matter in the interior of giant planets,
short-living dense plasmas produced by intense laser or heavy ion beams or
in explosive shock tubes, dusty (or complex) plasmas, and two-dimensional
electron systems on liquid helium \cite{Ichimaru_1982, Fortov_2005, 
Tahir_2006}. Therefore, it is a real challenge to study strongly coupled
EMP both theoretically and experimentally. 

In nonrelativistic EMP the interaction energy is given by the (screened) 
Coulomb potential. The Coulomb coupling parameter defined by
\be
\Gamma = \frac{q^2}{a \,T}
\label{coupling}
\ee
distinguishes between weakly coupled, $\Gamma \ll 1$, and strongly coupled
plasmas, $\Gamma {\buildrel > \over \sim} 1$. Here $q$ is the particle charge,
$a$ the interparticle distance and $T$ the kinetic temperature of the plasma 
component (electrons, ions, charged dust grains) under consideration. In the 
case of a degenerate plasma, e.g. the electron component in a white dwarf, 
the kinetic energy $T$ is replaced by the Fermi energy. Due to the strong 
interaction, the plasma can behave either as a gas or a liquid or even 
a solid (crystalline) system.

The case of an one-component plasma (OCP) with a pure Coulomb interaction
(a single species of charged particles in an uniform, neutralizing 
background) has been studied as a reference model for strongly coupled 
plasmas using simple models as well as numerical simulations in great 
detail \cite{Ichimaru_1982}. For $\Gamma > 172$ the plasma was shown to
form regular structures (Coulomb crystallization) \cite{Slattery_1980}. 
Below this critical value the OCP is in the supercritical state. For 
values of $\Gamma$ larger than about 50, it behaves like an ordinary 
liquid, while for small values below unity like a gas. Only if $\Gamma$ 
is large enough, the usual liquid behavior (Arrhenius law for the 
viscosity, Stokes-Einstein relation between self-diffusion and shear 
viscosity, etc.) appears due to caging of the particles (a single 
particle is trapped for some period of time in the cage formed by its 
nearest neighbors). For values of $\Gamma$, which are smaller than 
about 50, caging is not sufficiently strong and the system shows 
complicated, not yet understood transport properties. However, 
the short-range ordering typical for liquids shows up already 
for $\Gamma > 3$ \cite{Daligault_2006}. A gas-liquid transition, 
requiring a long-range attraction and a short-range repulsion, 
e.g. Lennard-Jones potential, does not exist in the OCP with particles 
of like-sign charges.

\begin{figure}
\epsfxsize17pc      
\centerline{\epsfbox{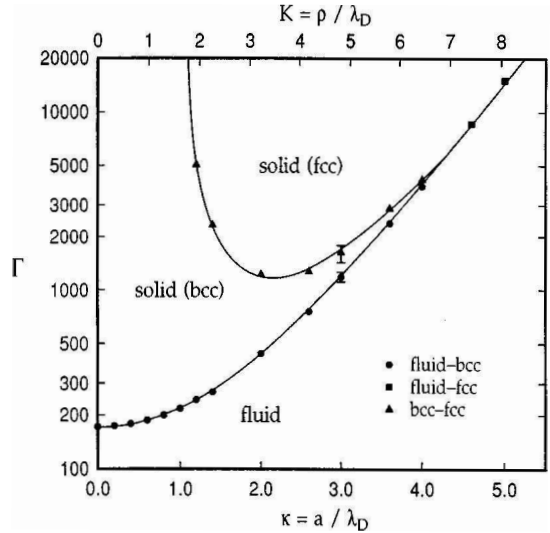}}
\caption{Phase diagram of a strongly coupled Yukawa system from 
Ref.~\cite{Hamaguchi_1997}.}
\label{phase}
\end{figure}

In realistic systems with a screened Coulomb interaction (Yukawa 
potential), the phase diagram can be shown in the 
$\Gamma$-$\kappa$-plane, where $\kappa = a/\lambda_D$ is the 
distance parameter with $\lambda_D$ being the Debye screening 
length. Numerical simulations based on molecular dynamics lead 
to the phase diagram shown in Figure~\ref{phase} \cite{Hamaguchi_1997}.

The first quantity of interest of sQGP is the coupling parameter. In 
analogy to nonrelativistic EMP, it is defined as \cite{Thoma_2005a}
\be
\Gamma = \frac{2Cg^2}{4\pi a \, T}= 1.5 - 5,
\label{coupling_QCD}
\ee
where $C$ is the Casimir invariant ($C=4/3$ for quarks and $C=3$ for gluons),
$a\simeq 0.5$ fm is the interparticle distance, and $T \simeq 200$ MeV is 
the QGP temperature corresponding to a strong coupling constant $g \simeq 2$.
The factor 2 in the numerator comes from taking into account the magnetic 
interaction in addition to the static electric (Coulomb) interaction, which 
are of the same magnitude in ultrarelativistic plasmas. The factor $4\pi$
in the denominator comes from using the Heavyside-Lorentz system in QCD 
as discussed in the Introduction. The distance parameter $\kappa $ of the 
QGP under the above conditions is rather small, typically between 1 and 3
\cite{Thoma_2005b}.

It should be noted, that we have assumed here a classical interaction 
potential corresponding to a one-gluon exchange. However, an effective 
potential taking into account higher order and non-perturbative effects 
may be much larger. This might be related to the fact that experimental 
data suggest a cross section enhancement for the parton interaction by 
more than an order of magnitude (see below). Hence the effective coupling parameter might be up to an order of magnitude larger than 
Equation~\ref{coupling_QCD}.

\begin{figure}\epsfxsize17pc
\epsfclipon              
\centerline{\epsfbox{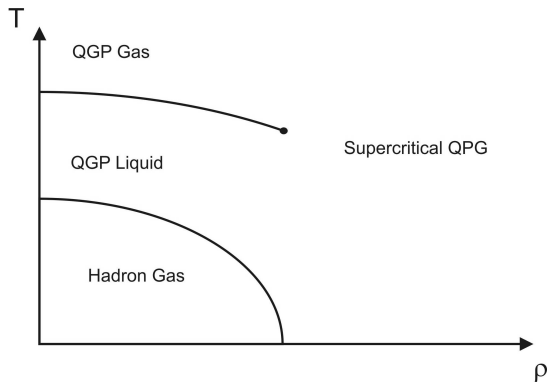}}
\caption{Sketch of a phase diagram of strongly interacting matter 
with a possible gas-liquid transition in the QGP phase. $T$ denotes
the temperature while $\rho$ is the baryon density.}
\label{gas_liquid}
\end{figure}

As discussed above, comparison to the OCP model as well as experimental 
data suggest that QGP close to the confinement phase transition
could be in a liquid phase. So, the question arises whether there is 
a phase transition from a liquid to a gaseous QGP, as sketched in 
Figure~\ref{gas_liquid}. For such a transition, a Lennard-Jones type 
interaction between the partons is required. However, the parton 
interaction in perturbative QCD is either purely repulsive or attractive
in the various interaction channels, e.g quark-antiquark or diquark 
channel. Due to non-linear effects caused by the strong coupling, 
however, attractive interactions can arise even in the case like-sign 
charges (see e.g. \cite{Tsytovich_2005}) leading to Lennard-Jones type 
potentials. Hence a gas-liquid transition in QGP with a critical point, 
proposed in \cite{Thoma_2006}, is not excluded and deserves further 
investigation.

An important quantity, which is very useful in theoretical and 
experimental studies of strongly coupled systems on the microscopic 
level, in particular in fluid physics \cite{Hansen_1986}, are
correlation functions. In particular, the pair correlation function 
and the static structure function provide valuable information on 
the equation of state of the system \cite{Hansen_1986}. The extension 
of the approach to the QGP has been proposed in \cite{Thoma_2005b}.

The static density-density autocorrelation function is defined 
for a classical system as \cite{Hansen_1986,Ichimaru_1982}
$$
G({\bf r}) = \frac{1}{N}\> \int d^3r' \> 
\langle \rho ({\bf r} + {\bf r'},t) \, \rho ({\bf r'},t) \rangle,
$$
where $N$ is the total number of particles and 
$$
\rho ({\bf r},t) = \sum_{i=1}^N 
\delta^{(3)}\big({\bf r}-{\bf r}_i(t)\big)
$$
is the local density of point particles with ${\bf r}_i(t)$ denoting 
the position of $i-$th particle at time $t$. The density-density 
autocorrelation function is related to the pair correlation function, 
which is defined as
$$
g({\bf r}) = \frac{1}{N}\> \langle \sum_{i,j,i\neq j}^N 
\delta^{(3)} ({\bf r}+{\bf r}_i-{\bf r}_j) \rangle,
$$
by the relation $G({\bf r}) = g({\bf r}) + \delta^{(3)} ({\bf r})$.
The static structure function, defined by 
$$
S({\bf p}) = \frac{1}{N}\> \langle \rho ({\bf p}) 
\rho (-{\bf p}) \rangle 
$$
with the Fourier transformed particle density, 
$$
\rho ({\bf p}) = \int d^3r \> \rho ({\bf r})\> 
e^{-i {\bf p} \cdot {\bf r}},
$$
is the Fourier transform of the density-density autocorrelation 
function
$$
S({\bf p}) = \int d^3r \> e^{-i {\bf p} \cdot {\bf r}}G({\bf r})\> .
$$

\begin{figure}
\epsfxsize17pc
\epsfclipon              
\centerline{\epsfbox{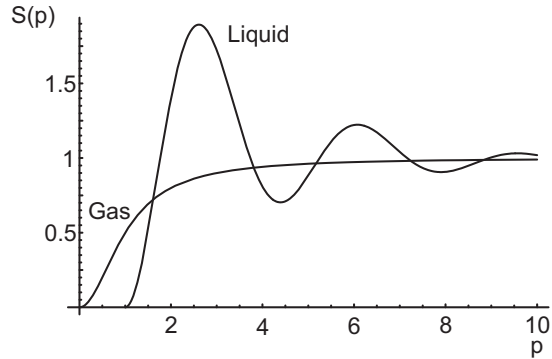}}
\caption{Sketch of the static structure functions vs. momentum 
in the gas and liquid phase in arbitrary units.}
\label{structure}
\end{figure}

The static structure function $ S({\bf p})$ is constant for 
${\bf p}\neq 0$ for uncorrelated particles \cite{Ichimaru_1973}. 
The typical behavior of the function in an interacting gas and in 
a liquid is sketched in Figure~\ref{structure}. The oscillatory 
behavior is caused by short range correlations, corresponding to 
a short range ordering typical for liquids. In the case of OCP, 
the oscillations appear for $\Gamma >3$ indicating a liquid behavior 
(albeit with non-standard transport properties) of the supercritical 
phase already for rather low values of $\Gamma $ \cite{Daligault_2006}.  

The static quark structure function can be related to the 
longitudinal gluon polarization tensor containing only the quark 
loop (Figure~\ref{polarization}) via
$$
S({\bf p}) = -\frac{12}{\pi g^2 n}\> \int_0^\infty d\omega \> 
\Im\Pi_L (\omega ,{\bf p})\>
\coth \frac{\omega}{2T},
$$
where $n=N/V=\langle \rho({\bf r})\rangle $ is the average particle 
density in a homogeneous system. As a reference for the strong coupling 
regime, the static quark structure function has been calculated in the 
weak coupling limit by resumming the polarization tensor in the 
high-temperature limit (Equation~\ref{polar-equi}) leading to 
\cite{Thoma_2005b}
\be
S({\bf p}) =\frac{2N_fT^3}{n}\> 
\frac{{\bf p}^2}{{\bf p}^2+m_D^2},
\label{e27}
\ee
where $m_D^2 = N_fg^2T^2/6$ is the quark contribution to the 
Debye screening mass. The static structure function given by 
Equation~\ref{e27} starts at zero for $|{\bf p}|=0$ and saturates 
at the uncorrelated structure function $S({\bf p})=2N_fT^3/n$ for 
large $|{\bf p}|$. Such a structure function  corresponds to an 
interacting Yukawa system in the gas phase (see Figure~\ref{structure}). 
Indeed, the pair correlation function, following from the Fourier 
transform of $S({\bf p})-1$, is
$$
g(r) = -\frac{N_fT^3}{2\pi n}\> \frac{m_D^2}{r}\> e^{-m_Dr},
$$
and it reproduces the Yukawa potential. 

In order to compute the structure function in strongly-coupled 
plasmas, molecular dynamics is used \cite{Ichimaru_1982}. Although 
QGP is not a classical system, as the thermal de Broglie wave length 
is of the same order as the interparticle distance, molecular 
dynamics might be useful as a first estimate \cite{Thoma_2006}. 
Using molecular dynamics for a classical sQGP 
\cite{Gelman_2006,Gelman:2006sr,Hartmann:2006nb}, the expected 
behavior described above has been qualitatively verified 
\cite{Gelman_2006}. In strongly coupled dense matter, where quantum 
effects are important, quantum molecular dynamics based on 
a combination of classical molecular dynamics with the density 
functional theory has been successfully applied \cite{Car_1985}. 
A generalization to the relativistic QGP has not been attempted 
so far. As an ultimate choice, lattice QCD could be used to 
calculate the structure or pair correlation functions. This would 
provide a test for the state of sQGP as well as the importance 
of quantum effects by comparing lattice results to classical 
molecular simulations.   

As a last application, we consider the influence of strong coupling 
on the cross sections entering transport coefficients (shear 
viscosity), stopping power, and other dynamical quantities of the 
plasma. Beside higher-order and non-perturbative quantum effects, 
there is already a cross section enhancement on the classical level.
The reason is that the Coulomb radius, defined as $r_c = q^2/E$ with 
the particle energy $E$, is of the order of the Debye screening 
length or $r_c$ is even larger than $\lambda_D$ in a strongly coupled 
plasma. Hence the standard Coulomb scattering formula has to be 
modified since the interaction with particles outside of the Debye sphere contributes significantly, and consequently the inverse 
screening length cannot be used as an infrared cutoff. This 
modification leads, for example, to the experimentally observed 
enhancement of the so-called ion drag force in complex plasmas 
which is caused by the ion-dust interaction \cite{Yaroshenko_2005}.

In the QGP at $T\simeq 200$ MeV, the ratio $r_c/\lambda_D$ equals 
1 - 5. It might enhance a parton cross section by a factor of 2 - 9 
\cite{Thoma_2005a} compared to perturbative results. An enhanced cross 
section reduces the mean free path $\lambda$, and consequently it 
reduces the viscosity $\eta$ as $\eta \sim \lambda$. An enhancement 
of the elastic parton cross section by more than an order of magnitude 
compared to perturbative results also explains the elliptic flow and 
particle spectra observed at RHIC \cite{Molnar_2002}. An infrared 
cutoff smaller than the Debye mass gives a natural explanation for 
this enhancement. We also note that if the cross section is enhanced, 
the collisional energy loss grows. On the other hand, the radiative 
energy loss is expected to be suppressed in the strongly coupled QGP 
by the Landau-Pomeranchuk-Migdal effect \cite{LPM}.

Finally, we mention two examples of strongly coupled systems which 
have not been considered in QGP physics but may be of relevance.
Strongly coupled plasmas such as two-dimensional Yukawa liquids 
\cite{Donko_2006} and dusty plasmas are non-Newtonian fluids, i.e. 
the shear viscosity depends on the shear rate (flow velocity) as 
it is well known in daily life from ketchup (shear thinning). The 
second example concerns nanofluidics. The expanding fireball in 
ultrarelativistic heavy-ion collisions has a transverse dimension 
of about 20 inter-particle distances (about 10 fm). Fluids consisting 
of such a low number of layers exhibit properties different from 
large fluid systems. For example, the shear flow does not show an 
continuous velocity gradient but jumps due to the adhesive forces 
between two layers. Such an behavior has been observed for example 
in complex plasmas.

\end{document}